\documentclass[]{scrartcl}
\usepackage[utf8]{inputenc}
\usepackage[T1]{fontenc}
\usepackage{lmodern}
\usepackage{amsmath}
\usepackage{enumitem}
\usepackage{amssymb}
\usepackage{graphicx}
\usepackage{hyperref} 
\usepackage[figure]{hypcap}
\usepackage[utf8]{inputenc}
\usepackage[T1]{fontenc}
\usepackage{lmodern}
\usepackage{mathtools}
\usepackage{mathrsfs}
\usepackage{amsmath}
\usepackage{graphicx}
\usepackage{amsthm}
\usepackage{svg}
\usepackage{placeins}
\usepackage{csquotes}
\usepackage{cite}

\DeclareMathOperator{\syn}{\mathrm syn}
\DeclareMathOperator{\ws}{\mathrm ws}

\usepackage[a4paper, left=2cm, right=2cm, top=2cm]{geometry}

\newtheorem{definition}{Definition}

\makeatletter
\def\moverlay{\mathpalette\mov@rlay}
\def\mov@rlay#1#2{\leavevmode\vtop{%
		\baselineskip\z@skip \lineskiplimit-\maxdimen
		\ialign{\hfil$\m@th#1##$\hfil\cr#2\crcr}}}
\newcommand{\charfusion}[3][\mathord]{
	#1{\ifx#1\mathop\vphantom{#2}\fi
		\mathpalette\mov@rlay{#2\cr#3}
	}
	\ifx#1\mathop\expandafter\displaylimits\fi}
\makeatother

\newcommand{\mba}{\mathbf{a}}
\newcommand{\mbb}{\mathbf{b}}

\title{From Babel to Boole: The Logical Organization of Information Decompositions}
\usepackage{authblk}

\author[1]{Aaron J. Gutknecht \thanks{Corresponding Author: agutkne@uni-goettingen.de}}
\author[1]{Abdullah Makkeh}
\author[1]{Michael Wibral}
\affil[1]{Campus Institute for Dynamics of Biological Networks, Georg-August University Göttingen, Germany}
\date{}

\begin{document}

\maketitle

\begin{abstract}
The conventional approach to the general Partial Information Decomposition (PID) problem has been redundancy-based: specifying a measure of redundant information between collections of source variables induces a PID via Moebius-Inversion over the so called redundancy lattice. Despite the prevalence of this method, there has been ongoing interest in examining the problem through the lens of different base-concepts of information, such as synergy, unique information, or union information. Yet, a comprehensive understanding of the logical organization of these different based-concepts and their associated PIDs remains elusive. In this work, we apply the mereological formulation of PID that we introduced in a recent paper to shed light on this problem. Within the mereological approach base-concepts can be expressed in terms of conditions phrased in formal logic on the specific parthood relations between the PID components and the different mutual information terms. We set forth a general pattern of these logical conditions of which all PID base-concepts in the literature are special cases and that also reveals novel base-concepts, in particular a concept we call ``vulnerable information''.
\end{abstract}

\section{Introduction} \label{sec:babel_to_boole:introduction}
Partial information decomposition (PID) is a powerful framework for dissecting the intricate relationships among multiple information sources and their joint contributions to a target variable. In the most simple case of two source variables $S_1$ and $S_2$ there is general agreement that the decomposition should contain four terms: the redundant information of $S_1$ and $S_2$ about $T$, the unique information of $S_1$ about $T$, the unique information of $S_2$ about $T$, and the synergistic information of $S_1$ and $S_2$ about $T$. There is also general agreement that these components should be related to the mutual information provided by subsets of these sources via the equations
\begin{align}
&I(S_1, S_2:T) = R + U_1 + U_2 + S\\ 
&I(S_1:T) = R + U_1 \\ 
&I(S_2:T) = R + U_2\;.
\end{align}
This system does not have a unique solution for the four components because we are short of one equation. A widely used approach to arrive at a determinate information decomposition is to fix one of the components and solve for the others using the above equation. The component most widely used for this purpose is the redundancy \cite{williams2010nonnegative, Harder2013, Griffith2014_common_randomness,Griffith_Ho2015,Barrett2015,ince2017measuring,finn2018pointwise,makkeh2021isx,Kolchinsky2022,magri2021shared,kleinman2021redundant,sigtermans2020pathbased}. But there are also some unique-information-based \cite{bertschinger2014quantifying,pakman2021estimating,james2018unique} and some  synergy-based approaches \cite{rosas2020operational,van2023pooling}. In principle, it is also possible to fix not an individual component but a certain combination of them, if this combination has an intuitive meaning. An example for this is the sum of the redundancy atom and the two unique information atoms. This describes the entirety of the information we can get from at least one information source and has been called \textit{union information}.  \cite{Griffith2014} and \cite{Kolchinsky2022} used this as the starting point to fix an information decomposition. We will refer to the information quantity  fixed in order to determine a full information decomposition as a PID \textit{base-concept}. In the general n-sources case, a base-concept will in fact encompass a whole set of quantities because the underlying system of equations becomes more and more undetermined. The key objective of this paper is provide a systematic study of PID base-concepts in this general case utilizing the mereological approach to partial information decomposition we introduced in \cite{gutknecht2021bits}.

There are three important reasons why this issue is of interest: First, there is a theoretical reason. Knowledge about the different possible ways to induce a PID provides insights into the structure of the problem. It makes clear which aspects of the original exposition of PID theory are essential and which aspects are replaceable. For example, does the concept of redundancy have a privileged role in PID theory? What about its underlying lattice structure? Furthermore, addressing the problem from the perspective of other base-concepts may also lead to constraints on possible solutions. For example, there have been numerous proposals for desirable properties or axioms on measures of redundant information \cite{Harder2013,Bertschinger2013_shared_info}, and also some for properties of synergistic information \cite{bertschinger2014quantifying}. A full account of synergy-based PID establishes a numerical connection between redundancy and synergy that allows us to determine whether the proposed properties of these two concepts are compatible. When they are not, the space of possible solutions to the problem is restricted.

The second reason pertains to the interpretation of information components. While it is true that in principle all base-concepts determine each other (fixing one, fixes all of them), the interpretation of measures that are not used as the base-concept will inevitably be that of a ``remainder''. Consider the case of two sources: if we specify their redundancy, then we can compute the unique information of each source by subtracting that redundancy from the total information provided by that source about the target. The synergy is then computed by subtracting redundancy and unique information from the total mutual information provided by both sources jointly, i.e. synergy is whatever remains if we subtract the other components from the total. This indirect definition makes the resulting notion of synergy quite intangible. By contrast, in a synergy-based PID, the synergy is directly defined in terms of the underlying joint distribution. This provides us with more control over its interpretation. Of course, the interpretational problem just described is shifted towards the non-synergistic components in this case. However, if in the application at hand synergy is of particular importance, a synergy-based decomposition might be preferable.

The third reason is a computational one. The number of distinct components in a PID grows super-exponentially with the number of information sources. Thus it becomes important to be able to compute useful summaries of the PID that do not require the computation of all components. An example for such a summary is the backbone decomposition introduced by \cite{rosas2020operational}. The components in this decomposition measure the information about the target for which access to exactly $k$ sources is required ($k = 1,\ldots,n)$. In this way the components provide a useful measure of the $k$-way interaction within the system of sources. The backbone components can be calculated very easily from a measure of synergy whereas it is not known how to compute them from a redundancy measure without having to compute all PID components. The same is true for the measure of ``representational complexity'' introduced in \cite{ehrlich2022partial}. Approaching the problem from the perspective of synergistic information makes this measure applicable to far larger networks since the computational cost scales only linearly with the number of sources in this case.

Our approach is as follows: In the next section, we review the mereological approach to PID using the example of redundant information and show how it expresses PID base-concepts in terms of their characteristic logical conditions on parthood relations. In Section \ref{sec:babel_to_boole:synergy_based_pid} we apply the approach to the construction of synergy-based PIDs. The analyses of redundancy-based and synergy-based PID naturally suggest a more general logical pattern of conditions for defining based-concepts which we will discuss in Section \ref{sec:babel_to_boole:logical_organization}. The resulting scheme comprises all base-concepts considered in the literature and also leads to new base-concepts. In particular a quantity we call ``vulnerable information'' and certain ``partner measures'' of the existing base-concepts which pick out the same information components but viewed from the perspective of different source collections. Section \ref{sec:babel_to_boole:lattices_axioms} addresses the implied properties of the different base-concepts as well as their associated lattices. Finally, in Section \ref{sec:babel_to_boole:rel_prev_appr} we discuss the relation of the ideas presented here to some previous approaches before presenting some general conclusions of our analysis in Section \ref{sec:babel_to_boole:conclusion}.

\section{The mereological approach to PID} \label{sec:babel_to_boole:mereological_approach}
In a recent paper we showed how to derive PID theory from considerations of parthood relations between information contributions \cite{gutknecht2021bits}. The key idea is that PID decomposes the information that the sources carry about the target into atomic contributions characterized by their parthood relations to the information provided by the different possible subsets of source variables. In other words, each information atom quantifies precisely that portion of the joint mutual information that stands in a particular constellation of parthood relationships to the $2^n$ 
different $I(\mathbf{a}:T)$ terms. Such constellations can be described by what we call \textit{parthood distributions}, i.e. Boolean functions ${f:\mathcal{P}(\{1,\ldots,n\}) \rightarrow \{0,1\}}$ that tell us for any subset $\mathbf{a} \subseteq \{1,\ldots,n\}$ of sources (referred to via their indices) whether the information atom described by $f$ is part of $I(\mathbf{a}:T)$. Parthood distributions form the cornerstone of mereological PID. Formally, they are defined as follows

\begin{definition}[Parthood Distribution]
A \textit{parthood distribution} is a function $f:\mathcal{P}(\{1,\ldots,n\}) \rightarrow \{0,1\}$ such that
\begin{enumerate}\label{babel_to_boole:axioms}
	\item $f(\emptyset) = 0$ ("There is no information in the empty set of sources")
	\item $f(\{1,\ldots,n\}) = 1$ ("All information is in the full set of sources")
	\item $\mathbf{a} \subseteq \mathbf{b} \text{ \& } f(\mathbf{a})=1 \rightarrow  f(\mathbf{b})=1$ ("All information in a set of sources is also in all of its supersets")
\end{enumerate}
\end{definition}

In a partial information decomposition there is one information atom $\Pi(f)$ per parthood distribution $f$. These considerations already tell us the intended meaning of the information atoms and how many atoms there are: one per parthood distribution. Since parthood distributions are formally non-constant, monotonic Boolean functions their number for $n$ sources is equal to the $n$-th Dedekind number minus two. Now, the question is: how many bits of information does each atom provide? In order to answer this question it is fruitful to think about how the atoms should be related to already known information quantities like mutual information. Given how the atoms are characterized it seems reasonable to demand that the following relation should be satisfied:
\begin{equation}\label{eq:babel_to_boole:consistency}
I(\mathbf{a}:T) = \sum\limits_{f(\mathbf{a})=1} \Pi(f) \hspace{0.5cm} (\textbf{consistency equation})
\end{equation}
This equation simply says that any mutual information should made up of all atoms which are part of it. And these are by construction all atoms $\Pi(f)$ such that $f(\mathbf{a})=1$. Summing over all such atoms will therefore yield the mutual information carried by the collection $\mathbf{a}$ about the target. We call Equation \ref{eq:babel_to_boole:consistency} the \textit{consistency equation} of PID. It provides constraints on quantitative solutions for the atoms $\Pi(f)$ by requiring them to be related to mutual information in a particular way. 

It is well known, however, that the consistency equation alone still leaves the problem severely under-constrained. We need some additional requirements to obtain a unique solution. This is traditionally achieved by invoking the concept of redundant information, which we generically denote by $I_\cap$. Based on the intended meaning of the atoms we should have
\begin{equation}\label{eq:babel_to_boole:atoms_redundancy}
I_{\cap}(\mathbf{a}_1,\ldots,\mathbf{a}_m:T) = \sum\limits_{\forall i f(\mathbf{a}_i)=1 } \Pi(f)
\end{equation}
In other words, the information shared by collections $\mathbf{a}_1,\ldots,\mathbf{a}_m$ about $T$ should consist of all information atoms that are part of \textit{each} of the $I(\mathbf{a}_i:T)$ contributions. But these are of course exactly those atoms $\Pi(f)$ such that $f(\mathbf{a})=1$ for \textit{all} $i=1,\ldots,m$. It can be shown that Equation \ref{eq:babel_to_boole:atoms_redundancy} is invertable so that once a measure of redundancy is specified a unique solution for the information atoms is implied. A PID obtained in this way is called a \textit{redundancy-based} PID. To see how this works, two insights are crucial.

First, note that Equation \ref{eq:babel_to_boole:atoms_redundancy} places a number of constraints on redundancy functions $I_\cap$:
\footnotesize
\begin{align} \label{eq:babel_to_boole:wb_axioms_redundancy_1}
     &\text{1. } I_\cap(\mathbf{a}_1,\ldots,\mathbf{a}_m:T) = I_\cap(\mathbf{a}_{\sigma(1)},\ldots,\mathbf{a}_{\sigma(m)}:T) \text{ for any permutation } \sigma \textbf{ (symmetry)} \\
     &\text{2. If } \mathbf{a}_i \supseteq \mathbf{a}_j \text{ for } i\neq j, \text{ then } I_\cap(\mathbf{a}_1,\ldots, ,\mathbf{a}_m:T) = I_\cap(\mathbf{a}_1,\ldots, \mathbf{a}_{i-1}, \mathbf{a}_{i+1},\ldots,\mathbf{a}_m:T)  \label{eq:babel_to_boole:wb_axioms_redundancy_3}\\ 
     & \hspace{8.5cm} \textbf{ (superset invariance)} \nonumber \\ 
    &\text{3. }I_\cap(\mathbf{a}:T) = I(\mathbf{a}:T) \textbf{ (self-redundancy)}
\end{align}

\normalsize
These constraints follow immediately from the properties of parthood distributions described above. In the literature they are known as the "Williams and Beer axioms" for redundancy functions since in their original exposition these properties play the role of axioms instead of being implied properties.  The first two of them, symmetry and superset invariance, imply that the domain of redundancy functions can be reduced to the set of antichains of the partial order $(\mathcal{P}(\{1,\ldots,n\}),\subseteq)$. We use the symbol $\mathcal{A}$ to denote the set of antichains without $\{\}$ and $\{\{\}\}$ since these do not correspond to any meaningful redundancy terms.

\begin{figure}[ht] 
	\centering
	\includegraphics[width=0.9\textwidth]{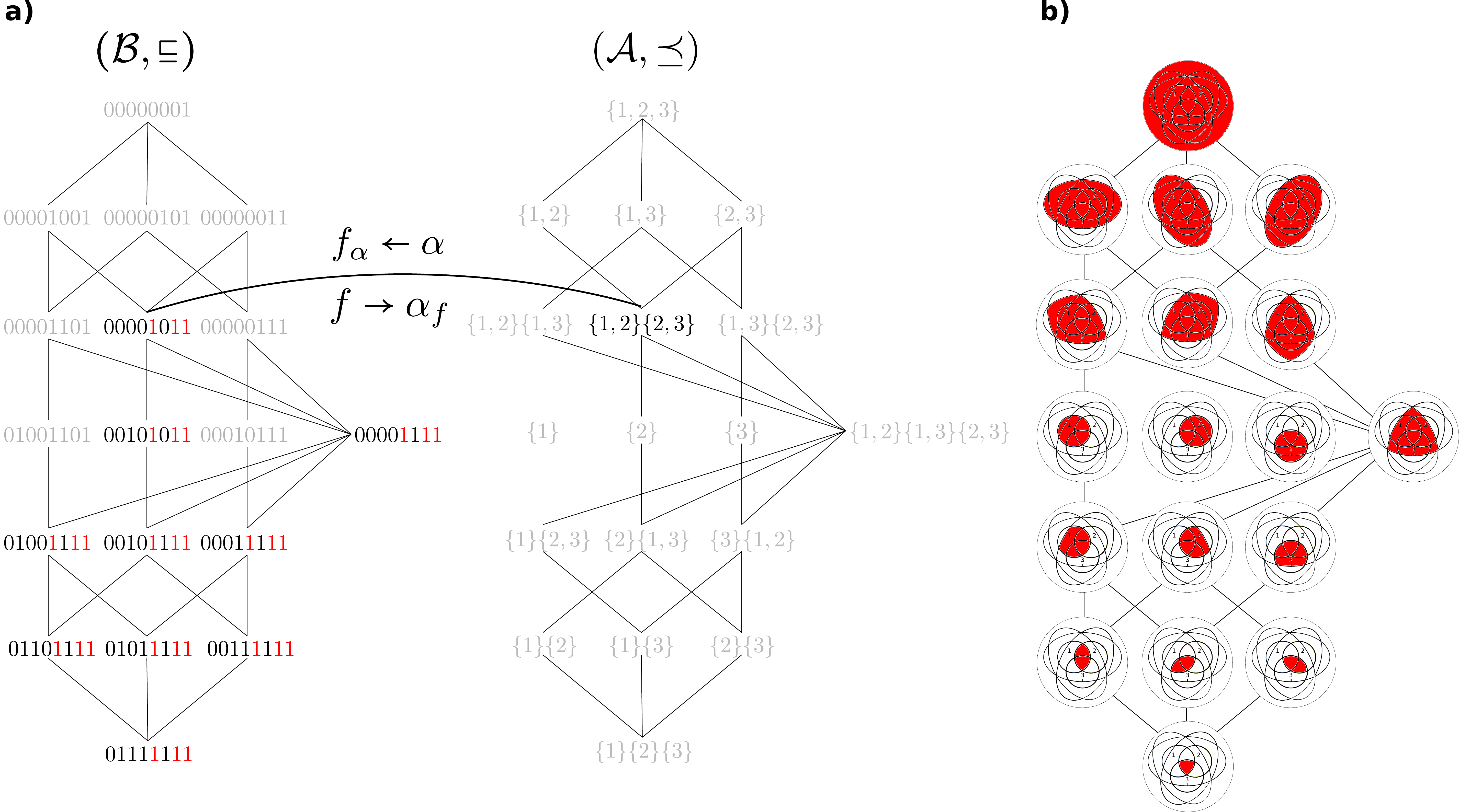}
	\caption{\textbf{a)} Parthood and redundancy lattices for $n=3$ sources. There is an isomorphism between the lattices such that the redundancy associated with a node in the redundancy lattice is equal to the sum of atoms associated with parthood distributions below and including the corresponding node in the parthood lattice. This is shown for the antichain $\{1,2\}\{2,3\}$. Note that we adhere to the standard convention of omitting the outermost brackets of the antichains. \textbf{b)} Information diagrams showing all possible redundancy terms and their nested structure.}
	\label{fig:babel_to_boole:3_redundancy_computation}
\end{figure}

The second important idea is that information atoms can be ordered quite naturally according to ``how easily'' they can be accessed. This can be expressed formally in terms of the following ordering on the parthood distributions:
\begin{equation}
f \sqsubseteq g  \Leftrightarrow \left(\forall~ \mathbf{a}\quad g(\mathbf{a})=1 \rightarrow f(\mathbf{a})=1\right)
\end{equation}
Intuitively, whenever the information described by $g$ is accessible via some collection, the information described by $f$ is also accessible via this collection. This order relation constitutes a lattice at the top of which we find the all-way synergy that can only be accessed if we know all sources and at the bottom of which we find the all-way redundancy that can be accessed via any source. Now this ordering stands in a close relationship to the concept of redundant information as expressed in Equation \ref{eq:babel_to_boole:atoms_redundancy}: consider an antichain $\alpha = \{\mathbf{a}_1,\ldots,\mathbf{a}_m\}$ and the parthood distribution that assigns the value one to \textit{exactly} these collections and their supersets. We denote this distribution by $f_\alpha$. We know that in general the redundant information $I_\cap(\mathbf{a}_1,\ldots,\mathbf{a}_m:T)$ is equal to the sum of all atoms with parthood distributions assigning the value one to all of the $\mathbf{a}_i$. But these atoms are, by construction, all atoms below and including $f_\alpha$ in the lattice. So we can rewrite Equation \ref{eq:babel_to_boole:atoms_redundancy} as
\begin{equation}\label{eq:babel_to_boole:moebius_inversion_red}
I_\cap(\alpha:T) = \sum\limits_{g \sqsubseteq f_\alpha} \Pi(g)
\end{equation}
The mapping $\alpha \rightarrow f_\alpha$ also induces a lattice structure on $\mathcal{A}$ that describes the nesting of redundant information terms. The induced ordering is
\begin{equation}
\alpha \preceq \beta \Leftrightarrow f_\alpha \sqsubseteq f_\beta
\end{equation}
The redundant information associated with an antichain $\alpha$ is included in any redundancy associated with antichains $\beta$ higher up in the lattice. The lattice $(\mathcal{A},\preceq)$ is the familiar \textit{redundancy lattice} intitally introduced by Williams and Beer \cite{williams2010nonnegative}. By construction the mapping $\alpha \rightarrow f_\alpha$ is an isomorphism between the redundancy lattice and the parthood lattice. The inverse is given by
\begin{equation}
f \rightarrow \alpha_f = \{\mathbf{a} \vert f(\mathbf{a})= 1 \text{ \& } \neg \exists \mathbf{b}\subset \mathbf{a} f(\mathbf{b})=1\}
\end{equation}
In other words, $\alpha_f$ is the set of minimal collections (with respect to $\subseteq$) that are assigned the value 1 by $f$. With these mappings one may also write Equation \ref{eq:babel_to_boole:atoms_redundancy} as a Moebius-Inversion over either $\mathcal{A}$ or $\mathcal{B}$ using the conventions $\Pi(\alpha) := \Pi(f_\alpha)$ and $I_\cap(f:T) := I_\cap(\alpha_f:T)$:
\begin{equation}
I_\cap(\alpha:T)= \sum\limits_{\beta\preceq \alpha} \Pi(\beta) \hspace{1cm} I_\cap(f:T)= \sum\limits_{g \sqsubseteq f} \Pi(g)
\end{equation}
These are now standard Moebius-Inversion formulas which are known to have a unique solution once a measure of redundant information $I_\cap$ is specified. This completes the redundancy-based PID story up to the choice of a concrete redundancy measure. Figure \ref{fig:babel_to_boole:3_redundancy_computation} illustrates the parthood and redundancy lattices as and how redundancy terms are expressed in terms of information atoms for the case $n=3$. In the next section, we apply the same mereological ideas in order to address the question of how a synergy-based PID can be constructed.

\FloatBarrier

\section{The construction of synergy based partial information decompositions} \label{sec:babel_to_boole:synergy_based_pid}
\subsection{Proper Synergy}
\begin{figure}[ht] 
	\centering
	\includegraphics[width=0.9\textwidth]{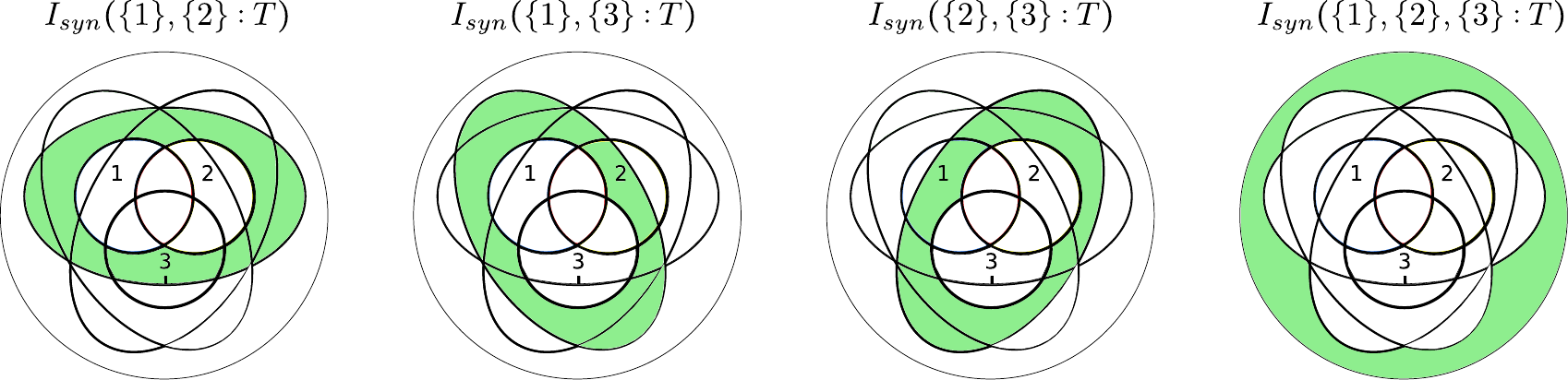}
	\caption{Mereological diagrams of the four independent synergy terms in the $n=3$ case. }
	\label{fig:babel_to_boole:proper_synergy_diagrams}
\end{figure}

Let us now apply the mereological ideas presented in the previous section to construct a synergy-based PID. To do so, the first question we have to ask is: can we in general express synergistic information $I_{\syn}$ as being made up out of certain information atoms $\Pi(f)$? Let us try to work out an answer. Intuitively, the synergy among collections $\mathbf{a}_1,\ldots,\mathbf{a}_m$ should certainly only contain information that is not contained in any individual collection $\mathbf{a}_i$. Otherwise, it would not make sense to call it synergistic. Translating this idea into a constraint on parthood distributions we can say that the synergy should only contain atoms $\Pi(f)$ such that $f(\mathbf{a}_i)=0$ for any i. Furthermore, it also seems reasonable that the synergy should not contain information that is accessible via some proper subset of sources contained in the $\mathbf{a}_i$. For instance, the synergistic information of sources $S_1$, $S_2$, and $S_3$ about the target should not already be contained in the combination of $S_1$ and $S_2$. Also, the synergy between $S_1$ and the combination $(S_2,S_3)$ should not be accessible if we only know $S_1$ and $S_2$. In terms of parthood distributions we can say the synergy should only contain information atoms $\Pi(f)$ such that $f(\mathbf{b})=0$ for all $\mathbf{b} \subset \bigcup \mathbf{a}_i$. This also includes the condition on individual $\mathbf{a}_i$ as a special case. We have now arrived at a negative constraint telling us which atoms are \textit{not} part of the synergy. So the only remaining question is which atoms \textit{are} part of it. Here it appears plausible to demand that if we had access to \textit{all} the collections $\mathbf{a}_i$, then we should obtain the synergistic information they carry about the target. As a parthood constraint this can be expressed as $f(\bigcup \mathbf{a}_i)=1$. Putting the negative and the positive constraint together this leads to the following relation between synergy $I_{\syn}$ and information atoms $\Pi$:
\begin{equation} \label{eq:babel_to_boole:proper_synergy}
I_{\syn}(\mathbf{a}_1,\ldots,\mathbf{a}_m:T) = \sum\limits_{\substack{\forall \mathbf{b} \subset \bigcup \mathbf{a}_i f(\mathbf{b}) = 0 \\ f(\bigcup \mathbf{a}_i)=1} } \Pi(f)    
\end{equation}
Now the crucial question is: can this relation be inverted to obtain a solution for all $\Pi(f)$ once a measure of synergy $I_{\syn}$ is provided? Unfortunately, the answer is no. The problem is that some of the equations coincide and hence the system is underdetermined. In fact, in the case of three sources, Equation \ref{eq:babel_to_boole:proper_synergy} only provides four constraints in addition to the consistency equation (11 would be needed). To see this, note first that given the relation above, $I_{\syn}$ has to be symmetric, idempotent, and invariant under \textit{subset} removal/addition. Hence, its domain can be reduced to the set of antichains. But there is a further constraint: whenever the union over two antichains is equal, the associated synergy must be equal. Formally,
{\small
\begin{equation}
\bigcup\mathbf{a}_i = \bigcup \mathbf{b}_j  \rightarrow I_{\syn}(\mathbf{a}_1,\ldots,\mathbf{a}_m:T) = I_{\syn}(\mathbf{b}_1,\ldots,\mathbf{b}_m:T) \textbf{ (Union Condition)}
\end{equation}
}
Accordingly, there can only be as many independent synergies as there are different non-empty unions (the synergy of the empty set has to be zero). Thus, we are left with seven synergy terms for $n=3$. Three terms correspond to the singletons $\{i\}$. For these, the condition in Equation \ref{eq:babel_to_boole:proper_synergy} reduces to $f(\mathbf{a})=1$ so that $I_{\syn}(\{i\}:T)=I(S_i:T)$. But this does not provide any constraint beyond the consistency equation. Three further terms correspond to the pairs of sources. And the final term corresponds to the full set of all three sources. It is only the last four terms that genuinely provide novel constraints on the information atoms. They are shown as mereological diagrams in Figure \ref{fig:babel_to_boole:proper_synergy_diagrams}.

In total, after defining a measure of synergy $I_{\syn}$ and given that we also have the consistency equation at our disposal, we are still short seven equations for $n=3$. An inversion of Equation \ref{eq:babel_to_boole:proper_synergy} is therefore not possible. Does this mean that there can be no such thing as a synergy-based PID? Not necessarily. It remains a possibility that there are alternative notions of synergistic information, notions that might still capture some, but necessarily not all, of the intuitive properties described above, and which allow for the required inversion. We will explore a minimal notion of synergy in the following section.

\subsection{Weak synergy} \label{sec:babel_to_boole:weak_synergy}
Let us strip our concept of synergistic information from anything but it's most essential property: the synergistic information carried by multiple collections of sources about the target should not be accessible via an individual collection $\mathbf{a}_i$. We will call the entirety of the information satisfying this condition the \textit{weak synergy} $I_{\ws}$ that collections $\mathbf{a}_1,\ldots,\mathbf{a}_m$ carry about the target \cite{gutknecht2021bits}.  Given this intended meaning of weak synergy it should stand in the following relation to the information atoms:
\begin{equation}
	I_{\ws}(\mathbf{a}_1,\ldots,\mathbf{a}_m:T) = \sum\limits_{\forall i f(\mathbf{a}_i)=0 } \Pi(f)
\end{equation}
In other words, we sum all atoms that are not part of any individual $I(\mba_i:T)$ contribution. Again, in order to determine whether this relation is invertible, we first ask which constraints on $I_{\ws}$ are implied by this condition. We obtain the following:
\footnotesize
\begin{align} \label{eq:babel_to_boole:wb_axioms_weak_synergy_1}
	 &\text{1. } I_{\ws}(\mathbf{a}_1,\ldots,\mathbf{a}_m:T) = I_{\ws}(\mathbf{a}_{\sigma(1)},\ldots,\mathbf{a}_{\sigma(m)}:T) \text{ for any permutation } \sigma \textbf{ (symmetry)}\\
	&\text{2. If } \mathbf{a}_i \subseteq \mathbf{a}_j \text{ for } i\neq j, \text{ then } {I_{\ws}(\mathbf{a}_1,\ldots,\mathbf{a}_m:T)} = {I_{\ws}(\mathbf{a}_1,\ldots,\mathbf{a}_{i-1},\mathbf{a}_{i+1},\ldots, \mathbf{a}_m:T)} \label{eq:babel_to_boole:wb_axioms_weak_synergy_3} \\
    &\hspace{8.9cm} \textbf{ (subset invariance)} \nonumber \\ 
	&\text{3. } I_{\ws}(\mathbf{a}:T) = I(\mathbf{a}^C:T|\mathbf{a}) \textbf{ (self-synergy)}
\end{align}
\normalsize
where $\mathbf{a}^C$ refers to the complement of $\mathbf{a}$. The first two conditions allow us to restrict the weak synergy to the set of antichains. Although this time we can exclude the antichains $\{\}$ and $\{\{1,\ldots,n\}\}$. The reason why the full set does not have to be included is that there is no information atom which is not contained in the information provided by the full set of sources. Accordingly, its weak synergy must be zero. Instead, the set containing the empty set $\{\{\}\}$ has to be included. The information not available if we do not know any source is of course all of the information in the sources. We refer to the domain of $I_{\ws}$ by $\mathcal{S}$. The system of self-synergy equations ensures that the resulting PID satisfies the consistency equation. This is because due to the chain rule for mutual information the conditions 
\begin{equation}
I(\mathbf{a}:T) = \sum\limits_{f(\mathbf{a})=1} \Pi(f) \text{ and } I(\mathbf{a^C}:T|\mathbf{a}) = \sum\limits_{f(\mathbf{a})=0} \Pi(f)
\end{equation}
are equivalent.

\begin{figure}[ht] 
	\centering
	\includegraphics[width=1\textwidth]{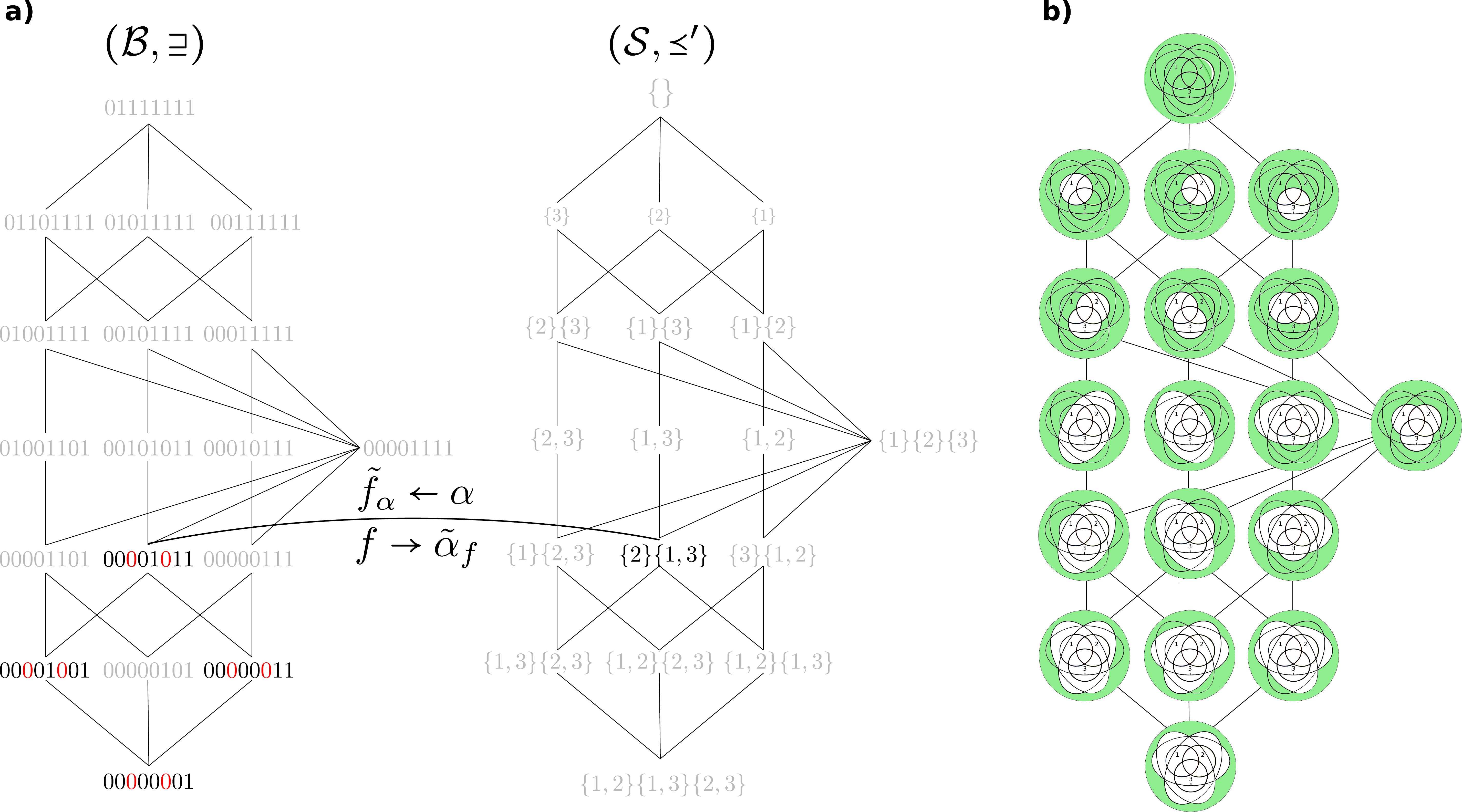}
	\caption{\textbf{a)} Parthood and synergy lattices for $n=3$ sources. There is an isomorphism between the lattices such that the weak synergy associated with a node in the synergy lattice is equal to the sum of atoms associated with parthood distributions above and including the corresponding node in the parthood lattice. This is shown for the antichain $\{2\}\{1,3\}$. Note that we adhere to the standard convention of omitting the outermost brackets of the antichains.  \textbf{b)} Mereological information diagrams depicting the different synergy terms.}
	\label{fig:babel_to_boole:3_synergy_computation}
\end{figure}

The relation between weak synergy and information atoms can be rewritten in terms of the ordering on parthood distributions. It is convenient to first turn this lattice upside-down so that the more easily accessible atoms are at the top, i.e. we are considering $(\mathcal{B},\sqsupseteq)$. By construction, the weak synergy of an antichain $\alpha = \{\mathbf{a}_1,\ldots, \mathbf{a}_m\}$ is equal to all atoms that such that $f(\mathbf{a}_i)=0$ for all $i=1,\ldots,m$. But these atoms are precisely the atoms associated with parthood distributions \textit{below and including} the parthood distribution ${\tilde{f_\alpha}}$ that assigns the value zero to \textit{exactly} all of the $\mathbf{a}_i$ and their subsets (in the upside-down parthood lattice): 
\begin{equation} \label{eq:babel_to_boole:relation_ws_atoms}
I_{\ws}(\alpha:T) = \sum\limits_{g\sqsupseteq {\tilde{f}_\alpha}} \Pi(g)
\end{equation}
All atoms further down in the ordering necessarily also assign the value zero to all $\mathbf{a}_i$ and additionally to some other collections as well (i.e. they are even harder to access). This computation is illustrated in Figure \ref{fig:babel_to_boole:3_synergy_computation}. What we can see from these considerations is that, just like the redundancies, the weak synergies are nested. The mapping $\alpha \rightarrow {\tilde{f_\alpha}}$ induces a lattice $(\mathcal{S}, \preceq^\prime)$ of antichains that describes this nesting. The ordering is given by
\begin{equation}
\alpha \preceq^\prime \beta \Leftrightarrow \tilde{f}_\alpha \sqsupseteq \tilde{g}_\beta
\end{equation}
We will refer to this lattice as the \textit{synergy lattice}. Weak synergies further down in this ordering are contained in synergies higher up. Just like the redundancy ordering, the synergy ordering on antichains also first appeared in a purely order-theoretic work \cite{Crampton2000} (written in a different but equivalent form). In the context of synergy-based PID it has been utilized by \cite{rosas2020operational} (as ``extended constraint lattice''), by \cite{chicharro2017synergy} (as ``information loss lattice''), and most recently by \cite{van2023pooling} (as "pooling-based lattice"). See Section \ref{sec:babel_to_boole:rel_prev_appr} for a discussion of the relation between these approaches and the mereological approach presented here.

By construction the mapping $\alpha \rightarrow \tilde{f}_\alpha$ is an isomorphism between the parthood lattice $(\mathcal{B},\sqsubseteq)$ and the synergy lattice $(\mathcal{S},\preceq^\prime)$. The inverse is given by
\begin{equation}
f \rightarrow \tilde{\alpha}_f = \{\mathbf{a}|f(\mathbf{a}=0) \text{ \& } \neg \exists \mathbf{b}\subset \mathbf{a} f(\mathbf{b})=0 \}    
\end{equation}
In other words, $ \tilde{\alpha}_f $ consists of the maximal sets $\mathbf{a}$ such that $f(\mathbf{a})=0$. Using this isomorphism and the conventions $I_{\ws}(f:T) := I_{\ws}(\tilde{\alpha}_f)$ and $\Pi(\alpha):= \Pi(\tilde{f}_\alpha)$ one can rewrite the relation between weak synergies and atoms as Moebius-Inversions over the parthood and synergy lattices respectively:
\begin{equation}\label{eq:babel_to_boole:weak_synergy_moebius}
I_{\ws}(f:T) = \sum\limits_{g\sqsupseteq f} \Pi(g) \hspace{2cm} I_{\ws}(\alpha:T) = \sum\limits_{\alpha \preceq^\prime \beta} \Pi(\beta)
\end{equation}
These relations can be inverted once a measure of weak-synergy is specified. We can see here that the construction of weak-synergy-based PIDs proceeds along the same lines as redundancy-based PID. The only difference is that the nesting of weak synergies is described by a different lattice structure. It is important to note that the intended interpretation of the information atoms $\Pi(f)$ remains exactly the same no matter if the PID is induced by a redundancy measure or a weak synergy measure. They still quantify the information that stands in the parthood relations described by $f$. 

Before we proceed to discuss how redundancy and weak synergy are special cases of a more general construction of PID base-concepts, we would like to consider an important interpretative point. Note that the formula on the right in \eqref{eq:babel_to_boole:weak_synergy_moebius} uses a different way to associate information atoms with antichains that the one used conventionally in the PID literature. In the standard way each information atom is associated with an antichain $\alpha$ in the redundancy lattice via the isomorphism $f \rightarrow \alpha_f$ we considered in Section \ref{sec:babel_to_boole:mereological_approach}. Given such an antichain $\alpha = \{\mathbf{a}_1,\ldots,\mathbf{a}_m\}$ the associated information atom $\Pi(\alpha)$ is the one which is part of the mutual information provided by any $\mathbf{a}_i$ and any superset thereof while it is not part of the mutual information provided by any other collection. In other words, the antichain tells us what the information atom is part of -- leaving it implicit what it is not part of. For instance, the atom $\Pi(\{1\})$ is the information uniquely contained in the first source. But there is also an alternative way that uses the synergy related isomorphism $f \rightarrow \tilde{\alpha}_f$, associating each atom with an antichain in the synergy lattice. Here the antichain tells us what the corresponding information atom is \textit{not} part of -- leaving it implicit what it is part of. In this interpretation the information atom $\tilde{\Pi}(\alpha)$ is not part of the mutual information provided by any $\mathbf{a}_i$ and any subset thereof while it is part of the information provided by any other collection. Accordingly, the unique information of the first source is $\tilde{\Pi}(\{1,\ldots,n\} \backslash \{1\})$ in this notation. It is of course straightforward to convert the two notations by composing the two mappings:
\begin{equation}
\tilde{\Pi}(\alpha) = \Pi(\beta_{\tilde{f}_\alpha}) \hspace{2cm} \Pi(\alpha) = \tilde{\Pi}(\tilde{\beta}_{f_\alpha})
\end{equation}
It is merely a matter of convenience which notation is used. However, when it comes to the interpretation of the information atoms it is important to be clear on this point. 

\section{The logical organization of PID base-concepts\label{sec:babel_to_boole:logical_organization}}
The construction of weak synergy and redundancy suggests a more general scheme for defining composite information measures. This construction defines the information associated with an antichain $\alpha$ in terms of sufficient, necessary, insufficient or unnecessary conditions on parthood or non-parthood with respect to either subsets or supersets of the $\mathbf{a} \in \alpha$. In the case of weak synergy, we are asking for all information such that it is a \textit{ sufficient condition } for an atom to be included in this information that \textit{it is not part} of the information provided by any \textit{subset} of the $\mathbf{a} \in \alpha$. We can rewrite the parthood condition of weak synergy (i.e.\ the condition $f$ has to satisfy so that $\Pi(f)$ is included in the weak synergy associated with $\alpha$) to make this more explicit. Setting $[n]=\{1,\ldots,n\}$:
\begin{equation}
 \forall~ \mathbf{b}\subseteq [n]: \exists~ \mathbf{a}\in \alpha~ \mathbf{b} \subseteq \mathbf{a} \rightarrow f(\mathbf{b}) = 0
\end{equation}
which picks out exactly the same information atoms for each $\alpha$ as the condition $\forall~\mathbf{a} \in \alpha:f(\mathbf{a})=0$. Similarly, in the case of redundant information we are asking for all information such that it is a \textit{sufficient condition} for an atom to be included in this information that  \textit{it is part} of the information provided by any \textit{superset} of the $\mathbf{a} \in \alpha$:
\begin{equation}
\forall~ \mathbf{b} \subseteq [n]: \exists~ \mathbf{a}\in \alpha~ \mathbf{b} \supseteq \mathbf{a} \rightarrow f(\mathbf{b}) = 1
\end{equation}
which picks out the same atoms as $\forall~ \mathbf{a} \in \alpha: f(\mathbf{a})=1$. In total the logical construction allows 16 possibilities. Before studying them in detail we would like to introduce a notion which will turn out to be very useful in the subsequent analysis.

\begin{definition}[Partner measure] Let $\mathcal{A^\ast}, \mathcal{A^{\ast\ast}}\subseteq \mathbb{A}$. Two information measures $I^{\ast}: \mathcal{A}^\ast \rightarrow \mathbb{R}$ and $I^{\ast\ast}: \mathcal{A}^{\ast\ast} \rightarrow \mathbb{R}$ are partner measures just in case there is a bijective mapping $\phi: \mathcal{A}^\ast \rightarrow \mathcal{A}^{\ast\ast}$ such that $I^\ast(\alpha:T)  = I^{\ast\ast}(\phi(\alpha):T) \forall \alpha \in \mathcal{A}^\ast$. 
\end{definition}
where $\mathbb{A}$ is the set of \textit{all} antichains of the partial order $([n],\subseteq)$, i.e. including both $\{\{\}\}$ and $\{\{1,\ldots,n\}\}$ as well as the empty set $\{\}$.   

\begin{figure}[] 
	\centering
	\includegraphics[width=1\textwidth]{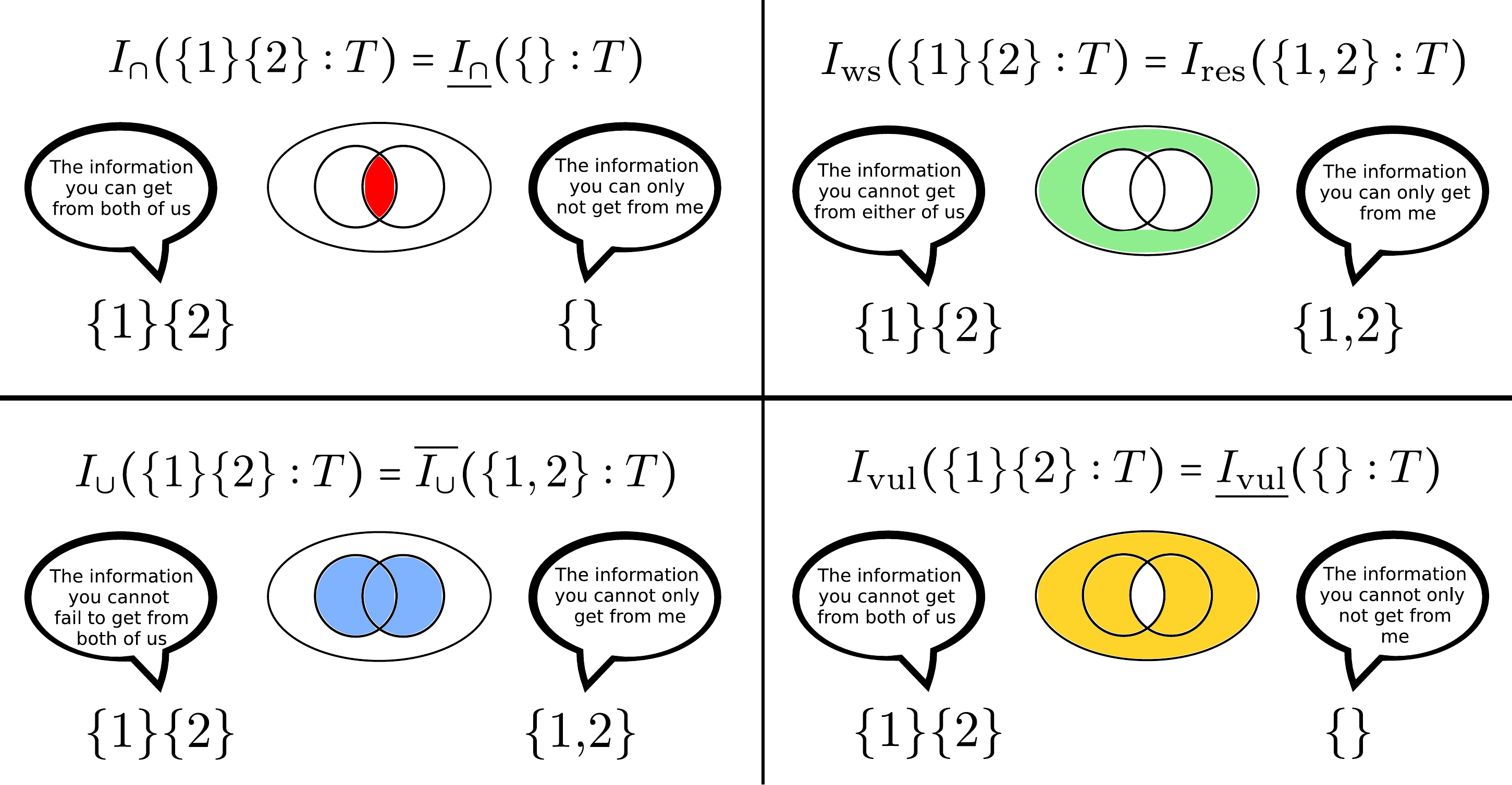}
	\caption{Intuitive interpretation of partner measures in the case $n=2$. \textit{Top left:} redundant information and its partner measure. The information which is redundant to both sources, $I_\cap(\{1\}\{2\}:T)$, is the information that we can only not get if we do not know any source, i.e. $\underline{I_\cap}(\{\}:T)$. \textit{Top right:} weak synergy and its partner measure. The information we cannot get from either source individually, $I_{ws}(\{1\}\{2\}:T)$, is the information we can only get if we know both sources at the same time, i.e. the information restricted to the full set of sources $I_{res}(\{1,2\}:T)=\overline{I_{ws}}(\{1,2\}:T)$. \textit{Bottom left:} union information and its partner measure. The union information, $I_\cup(\{1\},\{2\}:T)$, is the information we cannot fail to get from both individual sources. Or in other words, it is all information we can get from at least one individual source. This can equivalently be described as the information, ${\overline{I_\cup}(\{1,2\}:T)}$, we cannot \textit{only} get if we know both sources, i.e. for each component of the union information there is a way to access it that does not require full knowledge of both sources. \textit{Bottom right:} vulnerable information and its partner measure. The vulnerable information, $I_{vul}(\{1\}\{2\}:T)$, is all information we cannot get from both sources. This means that for each component of the vulnerable information there is a scenario in which we fail to obtain it \textit{other than the scenario in which we do not know any of the sources}. Therefore, it is the information we cannot only not get from the empty set of sources, i.e. $\underline{I_{vul}}(\{\}:T)$. }
	\label{fig:babel_to_boole:partner_measures}
\end{figure}

\begin{figure}[] 
	\centering
	\includegraphics[width=0.4\textwidth]{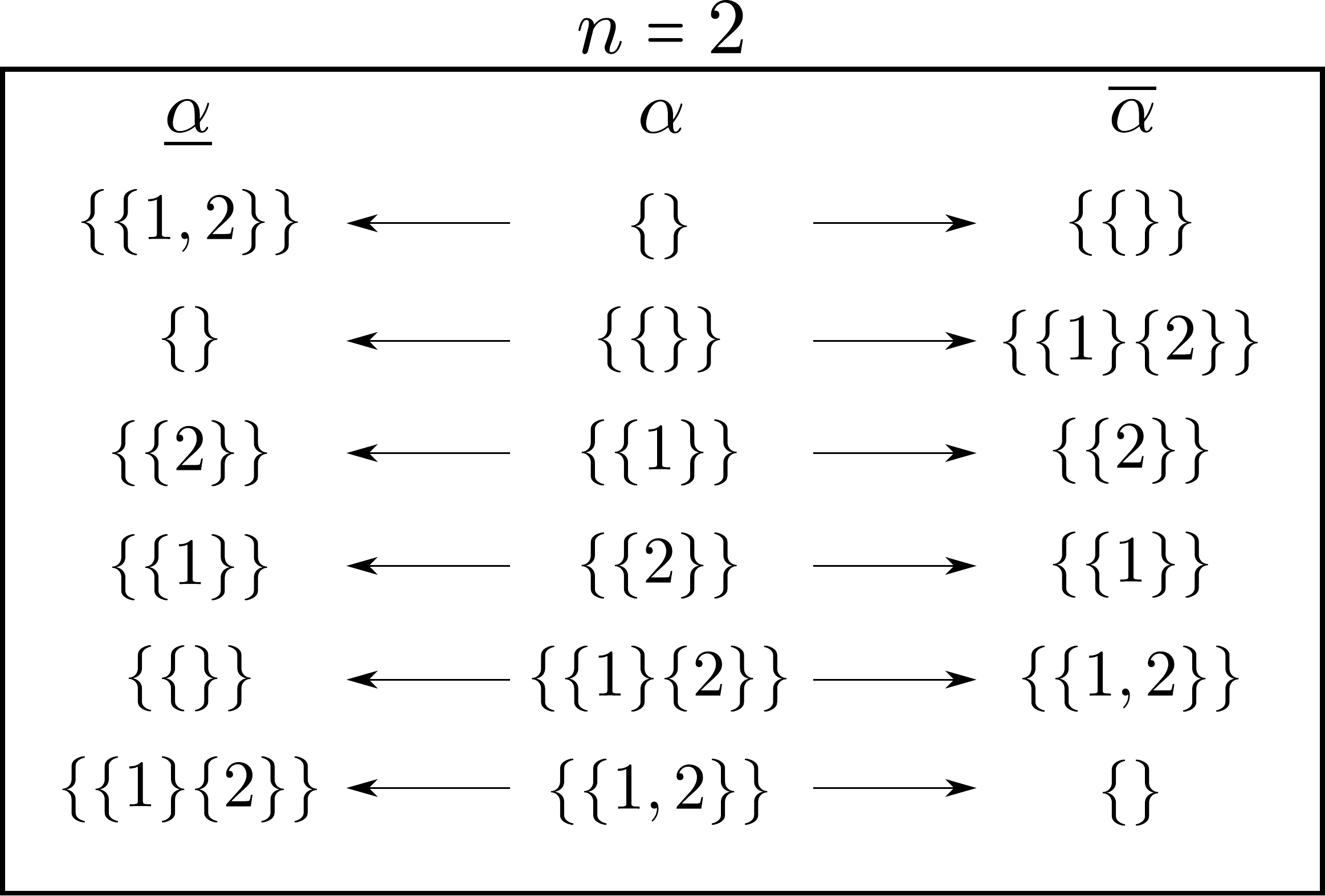}
	\caption{Mappings \ref{eq:babel_to_boole:alpha_overline}  and \ref{eq:babel_to_boole:alpha_underline} for $n=2$. Antichains $\alpha \in \mathcal{A}$ (middle column) are mapped to either $\underline{\alpha} \in \mathcal{A}$ (left column) or $\overline{\alpha} \in \mathcal{A}$ (right column).}
	\label{fig:babel_to_boole:partner_mapping_illustration}
\end{figure}

Partner measures quantify the same kind of information but viewed from the perspective of different collections. An example would be weak synergy and the "restricted information" we introduced in \cite{gutknecht2021bits}. The information we cannot get from any individual $\mba \in \alpha$ (weak synergy) is exactly the information we can \textit{only} get from other collections (i.e. the information restricted to these other collections), where the ``other'' collections are all non-subsets of the $\mba\in \alpha$. This is illustrated in the top right corner of Figure \ref{fig:babel_to_boole:partner_measures}. In the following, we will be interested specifically in partner measures with respect to the following two mappings between antichains
\begin{align} \label{eq:babel_to_boole:alpha_overline}
&\alpha \longmapsto \bar{\alpha} = \min\limits_{\subseteq} \left( \{\mathbf{b} \in [n] \mid \neg \exists~ \mathbf{a}\in \alpha: \mathbf{b}\subseteq \mathbf{a}\} \right)\\ \label{eq:babel_to_boole:alpha_underline}
&\alpha \longmapsto \underline{\alpha} =\max\limits_{\subseteq} \left( \{\mathbf{b} \in [n] \mid \neg \exists~ \mathbf{a}\in \alpha: \mathbf{b}\supseteq \mathbf{a}\} \right)
\end{align}
The first mapping collects the minimal non-subsets of the collections in $\alpha$ and the second one collects all the maximal non-supersets of these collections. Restricted information is a partner measure of weak synergy with respect to the first of the two mappings. Because the two mappings are inverses of each other (for proof see \ref{sec:babel_to_boole:proof_partner_inverses}), weak synergy is a partner measure of restricted information with respect to the second mapping. Figure \ref{fig:babel_to_boole:partner_mapping_illustration}
shows the two mappings for $n=2$.

Let us now consider all the possible cases of the general construction of information measures described above:
\paragraph{Sufficient Conditions} There are four conditions saying that being a subset/superset of some collection $\mathbf{a}\in \alpha$ is sufficient for parthood/non-parthood:
\begin{align}
&\forall \mathbf{b} \subseteq [n]: \exists \mathbf{a}\in \alpha~ \mathbf{b} \supseteq \mathbf{a} \rightarrow f(\mathbf{b}) = 1 
 &\forall \mathbf{b}\subseteq [n]: \exists \mathbf{a}\in \alpha~ \mathbf{b} \subseteq \mathbf{a} \rightarrow f(\mathbf{b}) = 0 \label{eq:babel_to_boole:sufficient_parthood} \\
&\forall \mathbf{b}\subseteq [n]: \exists\mathbf{a}\in \alpha~ \mathbf{b}\subseteq \mathbf{a} \rightarrow f(\mathbf{b})=1 &\forall \mathbf{b}\subseteq [n]: \exists\mathbf{a}\in \alpha~ \mathbf{b}\supseteq \mathbf{a} \rightarrow f(\mathbf{b})=0 \label{eq:babel_to_boole:sufficient_nonparthood}
\end{align}
We already discussed the first two conditions above. They correspond to redundancy and weak synergy respectively. The second two conditions are trivial. The first one because all parthood distributions satisfy $f(\{\})=0$. Thus, there is always a $\mathbf{b}$ for which the antecedent is true while the consequent is false. Accordingly, no information is included in the information described by the condition. Phrased differently, it is never sufficient for an information atom to be part of $I(\mathbf{b}:T)$ that $\mathbf{b}$ is a subset of some $\mathbf{a}\in \alpha$. Analogously, the second condition does not include any information atom because we always have $f(\{1,\ldots,n\}) = 1$. It is never sufficient for an information not to be part of $I(\mathbf{b}:T)$ that $\mathbf{b}$ is a superset of some $\mathbf{a}\in \alpha$.

\paragraph{Necessary Conditions} The following conditions express that being a subset/superset of some $\mathbf{a} \in \alpha $ is necessary for parthood/non-parthood:  
{\small
\begin{align}
&\forall\mathbf{b}\subseteq [n]: \neg\exists\mathbf{a}\in \alpha~ \mathbf{b}\supseteq \mathbf{a}   \rightarrow f(\mathbf{b})=1   &\forall \mathbf{b}\subseteq [n]:  \neg \exists\mathbf{a}\in \alpha~ \mathbf{b}\subseteq \mathbf{a}  \rightarrow f(\mathbf{b})=1  \label{eq:babel_to_boole:necessary_nonparthood} \\
&\forall \mathbf{b}\subseteq [n]: \neg \exists~\mathbf{a}\in \alpha~ \mathbf{b}\supseteq \mathbf{a} \rightarrow f(\mathbf{b})=0  &\forall \mathbf{b}\subseteq [n]:  \neg \exists\mathbf{a}\in \alpha~ \mathbf{b}\subseteq \mathbf{a} \rightarrow f(\mathbf{b})=0  \label{eq:babel_to_boole:necessary_parthood}
\end{align}
}
The first condition in \ref{eq:babel_to_boole:necessary_nonparthood} says that being a superset of some $\mathbf{a}\in \alpha$ is necessary for \textit{non-parthood}, i.e. in order for an atom to not be part of $I(\mathbf{b}:T)$ it must be the case that $\mathbf{b}$ is a superset of some $\mathbf{a}\in\alpha$. But this is never true. There is always a non-superset satisfying $f(\mathbf{b})=0$, namely $\mathbf{b}=\{\}$. Accordingly, the condition picks out none of the information atoms with the sole exception of the antichain $\alpha = \{\{\}\}$. Here it trivially picks out all atoms. 

The second condition in \ref{eq:babel_to_boole:necessary_nonparthood} is the partner measure of redundancy $\underline{I_\cap}$ because from the perspective of the $\underline{\mathbf{a}} \in \underline{\alpha}$ the non-subsets are exactly the $\mathbf{a} \in \alpha$ and their supersets so that $\underline{I_\cap}(\underline{\alpha}:T)= I_\cap(\alpha:T)$.  It says that being a subset of some $\mathbf{a}\in \alpha$ is necessary for non-parthood. Accordingly, the information picked out by the condition must be redundant with respect to all non-subsets of the collections at which $\underline{I_\cap}$ is evaluated. For an illustration see the top left corner of Figure \ref{fig:babel_to_boole:partner_measures}.

In \ref{eq:babel_to_boole:necessary_parthood}, the first condition describes the partner measure of weak synergy $\overline{I_{\text{ws}}}$ because from the perspective of the $\overline{\mathbf{a}} \in \overline{\alpha}$ the non-supersets are exactly the $\mathbf{a}\in \alpha$ and their subsets so that ${\overline{I_\text{ws}}(\overline{\alpha}:T)= I_\text{ws}(\alpha:T)}$. It says that being a superset of some $\mathbf{a}\in \alpha$ is necessary for \textit{parthood}, i.e. it captures information that may only be contained in the collections at which it is evaluated and their supersets. This is the \textit{restricted information} discussed above (see also top right corner of Figure \ref{fig:babel_to_boole:partner_measures}).

The second condition in \ref{eq:babel_to_boole:necessary_parthood} says that being a subset of some $\mathbf{a}\in \alpha$ is necessary for \textit{parthood}. But this is never the case.  There is always a non-subset satisfying $f(\mathbf{b})=1$, namely $\mathbf{b}=\{1,\ldots,n\}$. Accordingly, the condition picks out none of the information atoms with the sole exception of the antichain $\alpha = \{\{1,\ldots,n\}\}$ where it trivially picks out all atoms. 

\paragraph{Insufficient Conditions} The conditions expressing that being a subset/superset of some $\mathbf{a}\in\alpha$ is insufficient for parthood/non-parthood are 
{\small
\begin{align}
&\neg(\forall \mathbf{b}\subseteq [n]: \exists\mathbf{a}\in \alpha~ \mathbf{b}\supseteq \mathbf{a} \rightarrow f(\mathbf{b})=1)  &\neg(\forall \mathbf{b}\subseteq [n]: \exists\mathbf{a}\in \alpha~ \mathbf{b}\subseteq \mathbf{a} \rightarrow f(\mathbf{b})=1)\label{eq:babel_to_boole:insuff_for_parthood} \\
&\neg(\forall\mathbf{b}\subseteq[n]: \exists\mathbf{a}\in \alpha~ \mathbf{b}\supseteq \mathbf{a} \rightarrow f(\mathbf{b})=0) &\neg(\forall \mathbf{b}\subseteq [n]: \exists\mathbf{a}\in \alpha~ \mathbf{b}\subseteq \mathbf{a} \rightarrow f(\mathbf{b})=0) \label{eq:babel_to_boole:insuff_for_nonparthood}
\end{align}
}
The first condition in \eqref{eq:babel_to_boole:insuff_for_parthood} leads to a measure of information that has not been described in the literature before. Intuitively, it describes the ``the information we do not get from at least one $\mathbf{a} \in \alpha$''. One might call this \textit{vulnerable information} because it is not completely redundant with respect to the $\mba \in \alpha$ and hence may be lost if we loose access to some of these collections (or is not contained in any of them in the first place). It is the complement of the redundancy. The second condition  in \eqref{eq:babel_to_boole:insuff_for_parthood} is trivial. It includes all atoms because there is always a subset of the $\mbb \in  \alpha$ for which $f(\mbb) = 0$, namely $\mbb = \{\}$. Similarly, the first condition in \eqref{eq:babel_to_boole:insuff_for_nonparthood} includes all atoms because there is always a superset of the $\mbb \in  \alpha$ for which $f(\mbb) = 1$, namely $\mbb = \{1,\ldots,n\}$. The second condition in \eqref{eq:babel_to_boole:insuff_for_nonparthood} describes the union information, i.e. the information we can obtain from at least one $\mba \in \alpha$. 
\paragraph{Unnecessary Conditions} Finally, there are four conditions saying that being a subset/superset of some $\mathbf{a}\in \alpha$ is unnecessary for parthood/non-parthood:
{\small
\begin{align}
&\hspace{-1.5em}\neg(\forall\mathbf{b}\subseteq [n]: \neg\exists\mathbf{a}\in \alpha~ \mathbf{b}\supseteq \mathbf{a}   \rightarrow f(\mathbf{b})=1 )  &\neg(\forall \mathbf{b}\subseteq [n]:  \neg \exists\mathbf{a}\in \alpha~ \mathbf{b}\subseteq \mathbf{a}  \rightarrow f(\mathbf{b})=1)  \label{eq:babel_to_boole:unnec_for_nonparthood} \\
&\hspace{-1.5em}\neg(\forall \mathbf{b}\subseteq [n]: \neg \exists~\mathbf{a}\in \alpha~ \mathbf{b}\supseteq \mathbf{a} \rightarrow f(\mathbf{b})=0  )  &\neg(\forall \mathbf{b}\subseteq [n]:  \neg \exists\mathbf{a}\in \alpha~ \mathbf{b}\subseteq \mathbf{a} \rightarrow f(\mathbf{b})=0  )\label{eq:babel_to_boole:unnec_for_parthood}
\end{align}
}
The first condition in \ref{eq:babel_to_boole:unnec_for_nonparthood} is trivial. It says that being a superset of some $\mathbf{a} \in \alpha$ is not necessary for non-parthood. But this is true for all antichains and information atoms because there always a non-superset for which $(f(\mathbf{b}))=0$, namely $\mathbf{b}=\{\}$. The only exception is $\alpha = \{\{\}\}$ for which the condition trivially picks out no information atom. The second condition in \ref{eq:babel_to_boole:unnec_for_nonparthood} is the partner measure $\underline{I_{\text{vul}}}$ of vulnerable information because from the perspective of $\underline{\alpha}$ the non-subsets are exactly the $\mathbf{a}\in \alpha$ and their supersets so that $ \underline{I_{\text{vul}}}(\underline{\alpha}:T)= I_\text{vul}(\alpha:T)$. For an intuitive description of vulnerable information and its partner measure see the bottom right corner of Figure \ref{fig:babel_to_boole:partner_measures}.

The first condition in \ref{eq:babel_to_boole:unnec_for_parthood} is the partner measure $\overline{I_\cup}$ of union information because from the perspective of $\overline{\alpha}$ the non-supersets are exactly the $\mathbf{a}\in\alpha$ and their subsets so that $\overline{I_\cup}(\overline{\alpha}:T)= I_\cup(\alpha:T)$. For an intuitive description of union information and its partner measure see the bottom left corner of Figure \ref{fig:babel_to_boole:partner_measures}. The second condition in \ref{eq:babel_to_boole:unnec_for_parthood} is trivial because it says that being a subset of some $\mathbf{a}\in\alpha$ is not necessary for parthood. But this is true for all antichains and information atoms since there is always a non-subset for which $f(\mathbf{b})=1$, namely $\mathbf{b}=\{1,\ldots,n\}$. The only exception is $\alpha = \{\{1,\ldots,n\}\}$ where the condition trivially picks out no atom. 

\begin{figure}[ht] 
	\centering
	\includegraphics[width=1\textwidth]{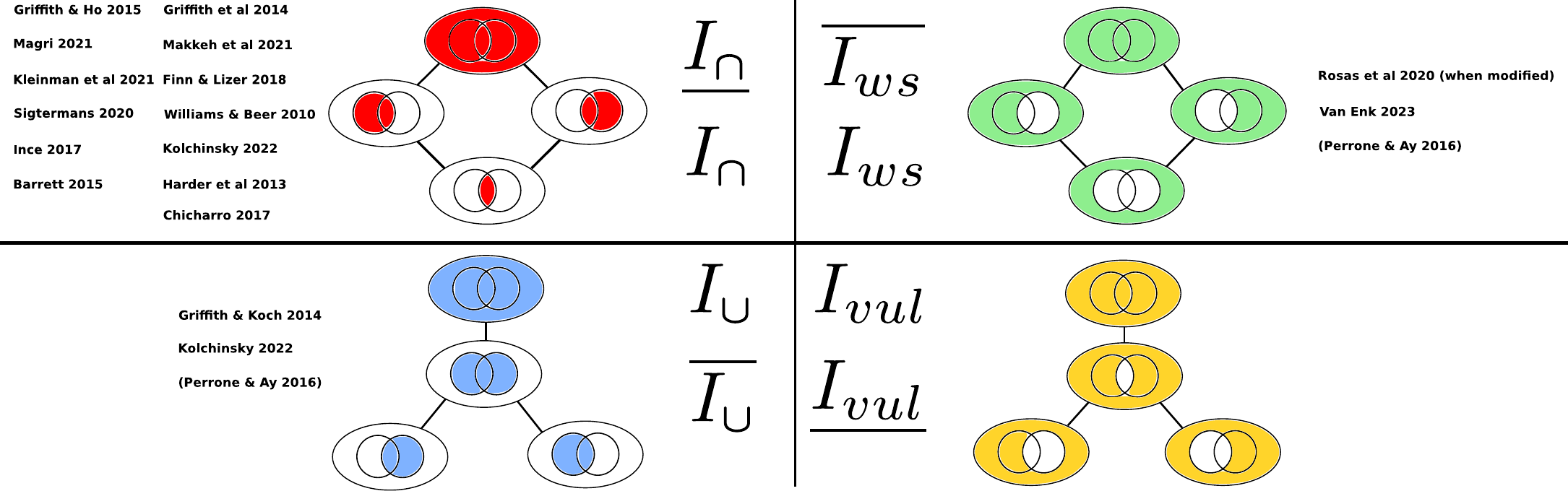}
	\caption{Scheme of four equivalence classes of partner measures. Previous PID approaches are categorized in the appropriate quadrants.}
	\label{fig:babel_to_boole:logical_scheme}
\end{figure}

\paragraph{} So in total we obtain four pairs of partner measures as shown in Figure \ref{fig:babel_to_boole:logical_scheme} for the case $n=2$. The Figure also locates previous PID approaches within this scheme. Thus far, there has been no proposal utilizing vulnerable information as a PID base-concept. Furthermore, all proposals in the literature are based on $I_\cap$, $I_\cup$, or $I_{ws}$ rather than their partner measures. Some comments are in order in particular about the weak synergy quadrant: The measure of "synergistic disclosure" by Rosas et al \cite{rosas2020operational} is very close in spirit to what we have called weak synergy here but only leads to a standard PID when it is modified appropriately. This is discussed in Section \ref{sec:babel_to_boole:modified_sd} below. The approach by Perrone \& Ay \cite{perrone2016hierarchical} does not attempt to construct a PID but rather a decomposition of joint mutual information into interactions of orders 1 to n. In the two-sources case this amounts to defining union information and the synergy atom which is why we included it in parentheses.

\paragraph{}Before discussing the implied properties and associated lattices of the different base-concepts we would like to briefly address a base-concept that we have not considered so far: unique information. This has only been utilized in the two-sources case \cite{bertschinger2014quantifying,pakman2021estimating,james2018unique}. However, we argued in \cite{gutknecht2021bits} that one may generalize the concept so that it becomes a base-concept in the general case as well. Given collections of source variables $\alpha$ one may think of the unique information associated with these collections as "the information contained in all of the $\mathbf{a} \in \alpha$ but nowhere else". In other words, it consists of the information atoms $\Pi(f)$ where $f(\mathbf{b})=1$ if $\mathbf{b} \supseteq \mathbf{a}$ for some $\mathbf{a}\in \alpha$ and $f(\mathbf{b})=0$ otherwise. There is only one such information atom, namely the atom $\Pi(f_\alpha)$ so that we have $I_\text{unq}(\alpha:T) = \Pi(f_\alpha) = \Pi(\alpha)$ (see Section \ref{sec:babel_to_boole:weak_synergy} above for an explanation of this notation). Hence, a unique information based PID amounts to defining the information atoms directly. Unique information can also be described by a logical condition similar to the ones we discussed above. It is captured by a \textit{sufficient and necessary} condition with respect to parthood in supersets of the $\mathbf{a}_i$:
{\small
\begin{equation}
 \forall\mathbf{b} \subseteq [n]: \exists \mathbf{a}\in \alpha \mathbf{b} \supseteq \mathbf{a} \rightarrow f(\mathbf{b}) = 1 \text{ \& }  \forall \mathbf{b}\subseteq [n]: \neg \exists~\mathbf{a}\in \alpha~ \mathbf{b}\supseteq \mathbf{a} \rightarrow f(\mathbf{b})=0 
\end{equation}
}
This condition is the logical conjunction of the conditions for $I_\cap$ and $\overline{I_\text{ws}}$. It also has a natural partner measure arising from the conjunction of the $I_\text{ws}$ and $\underline{I_\cap}$ conditions which amounts to a \textit{sufficient and necessary} condition with respect to non-parthood in subsets of the $\mathbf{a}_i$:
{\small
\begin{equation}
\forall \mathbf{b}\subseteq [n]: \exists \mathbf{a}\in \alpha~ \mathbf{b} \subseteq \mathbf{a} \rightarrow f(\mathbf{b}) = 0 \text{ \& } \forall \mathbf{b}\subseteq [n]:  \neg \exists\mathbf{a}\in \alpha~ \mathbf{b}\subseteq \mathbf{a}  \rightarrow f(\mathbf{b})=1
\end{equation}
}
This describes the partner measure $\underline{I_\text{unq}}$ and can be interpreted as "the information we do not get from any of the $\mathbf{a}\in \alpha$ but anywhere else". It satisfies $\underline{I_\text{unq}}(\alpha:T) = \Pi(\tilde{f}_\alpha) = \tilde{\Pi}(\alpha)$ (see Section \ref{sec:babel_to_boole:weak_synergy} above for an explanation of this notation).

\FloatBarrier

\section{Properties and Lattices} \label{sec:babel_to_boole:lattices_axioms}
Each of the information measures discussed in the previous section is associated with a particular lattice (or semi-lattice) describing its nested structure (except of course unique information since it is not nested and simply has to satisfy the consistency equation \ref{eq:babel_to_boole:consistency}). For redundancy and weak synergy these are the lattices $(\mathcal{A},\preceq)$ and $(\mathcal{S},\preceq^\prime)$ as introduced in Section \ref{sec:babel_to_boole:mereological_approach} and \ref{sec:babel_to_boole:synergy_based_pid}. Furthermore, each information measure has a range of fundamental properties following from their characteristic parthood conditions. For redundancy and weak synergy these are the above Equations \ref{eq:babel_to_boole:wb_axioms_redundancy_1}-\ref{eq:babel_to_boole:wb_axioms_redundancy_3} and \ref{eq:babel_to_boole:wb_axioms_weak_synergy_1}-\ref{eq:babel_to_boole:wb_axioms_weak_synergy_3}, respectively. The corresponding lattices and properties of the other base-concepts can be derived easily utilizing their relations to redundancy and weak synergy as well as the mappings $\alpha \rightarrow \overline{\alpha}$ and $\alpha \rightarrow \underline{\alpha}$.

\paragraph{The redundancy partner $\underline{I_\cap}$:} The domain of the $\underline{I_\cap}$ is the image of $\mathcal{A}$ under $\alpha \rightarrow \underline{\alpha}$, i.e. $\underline{\mathcal{A}} = \mathbb{A} \backslash \{\{\{1,\ldots,n\}\}, \{\}\} = \mathcal{S}$. In order to find the ordering relation note that
\begin{equation}
\underline{I_\cap}(\alpha:T) = \sum\limits_{g \sqsubseteq \tilde{f}_\alpha} \Pi(g)
\end{equation}
The left hand side expresses the information at most not contained in the $\mathbf{a}\in \alpha$ and their subsets. But this is equal to the information atom $\Pi(\tilde{f}_\alpha)$, which is not contained \textit{exactly} in all $\mathbf{a} \in \alpha$ and subsets thereof, plus all information atoms further down the parthood lattice, i.e. all more accessible atoms. Hence, the appropriate ordering relation is the inverted weak synergy ordering (compare Equation \ref{eq:babel_to_boole:relation_ws_atoms} above)  so that the lattice for $\underline{I_\cap}$ is $(\mathcal{S},\succeq^\prime)$. In other words, using  $\underline{I_\cap}$ as a base-concept amounts to performing an upwards Moebius-Inversion over the synergy lattice. $\underline{I_\cap}$ is symmetric, subset-invariant and satisfies the condition
\begin{equation}
\underline{I_\cap}([n]\backslash\{i_1\},\ldots,[n]\backslash\{i_m\}:T) = I(\{i_1,\ldots,i_m\}:T)
\end{equation}

\paragraph{The weak synergy partner $\overline{I_\text{ws}} = I_\text{res}$:} Analogously, the domain of  $\overline{I_\text{ws}}$ is the image of $\mathcal{S}$ under $\alpha \rightarrow \overline{\alpha}$, i.e. $\overline{\mathcal{S}} = \mathbb{A} \backslash \{ \{\{\}\}, \{\} \} = \mathcal{A}$, equipped with the inverted redundancy ordering because
\begin{equation}
\overline{I_\text{ws}}(\alpha:T) = \sum\limits_{g \sqsupseteq f_\alpha} \Pi(g)
\end{equation}
The information at most contained in a superset of the $\mathbf{a} \in \alpha$ is equal to the information atom $\Pi(f_\alpha)$ which is contained \textit{exactly} in all $\mathbf{a} \in \alpha$ and their supersets, plus all information atoms further down the parthood lattice, i.e. all even less accessible atoms. Hence, the nesting is described by the lattice $(\mathcal{A},\succeq)$. In other words, using  $\overline{I_\text{ws}}$ as a base-concept amounts to performing an upwards Moebius-Inversion over the redundancy lattice. $\overline{I_\text{ws}}$ is symmetric, superset-invariant and satisfies the condition
\begin{equation}
\overline{I_\text{ws}}(\{i_1\},\ldots,\{i_m\}:T) = I(\{i_1,\ldots,i_m\}:T|\{i_1,\ldots,i_m\}^C)
\end{equation}

\begin{figure}[ht] 
	\centering
	\includegraphics[width=0.75\textwidth]{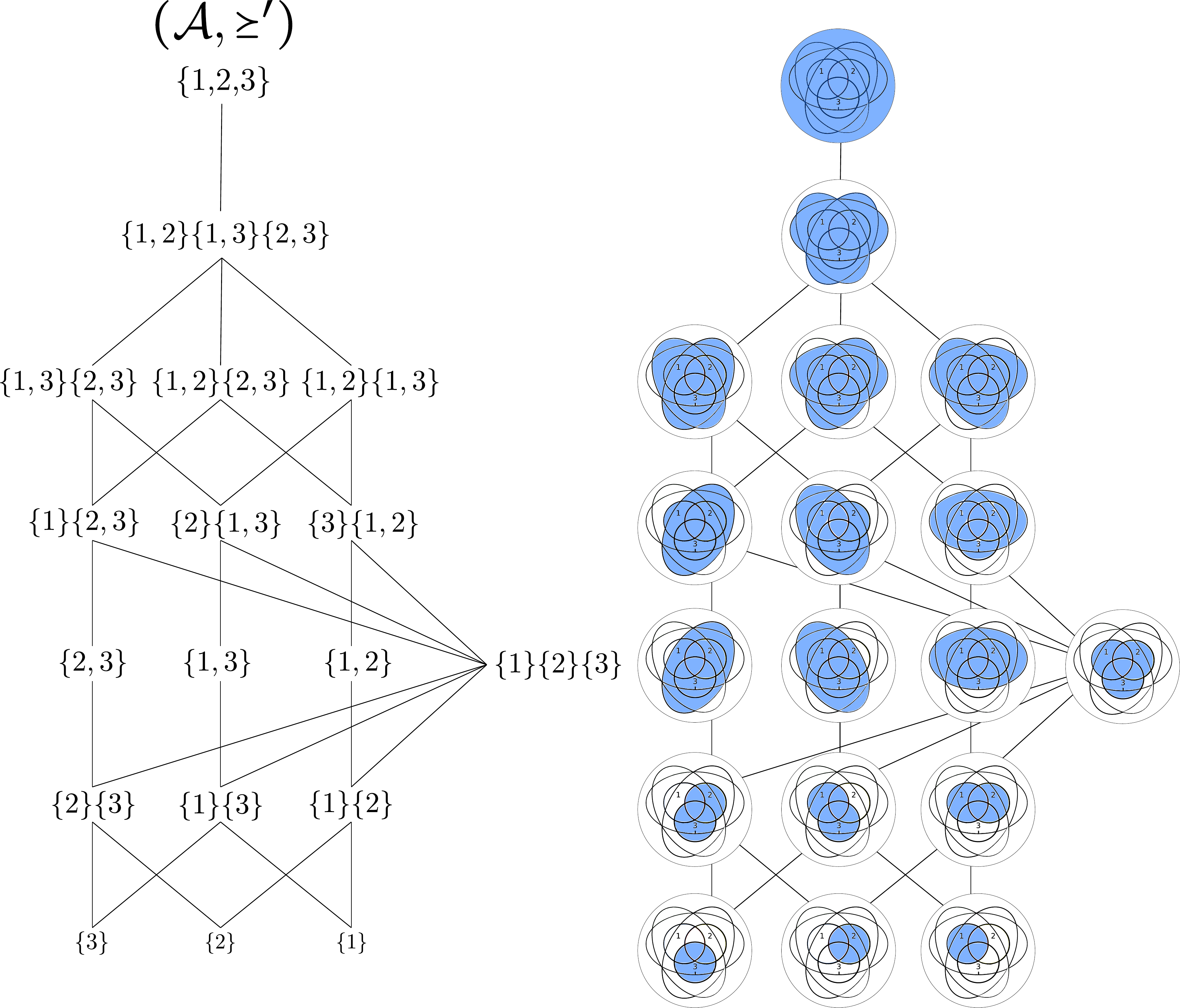}
	\caption{Left: Union information semi-lattice for $n=3$ sources. Right: Mereological information diagrams depicting the different union information terms.}
	\label{fig:babel_to_boole:3_union_lattice}
\end{figure}

\paragraph{Union information and its partner:} Since union information is the complement of weak synergy, i.e. the atoms summed over to obtain the union information are exactly the atoms not summed over to obtain the weak synergy and vice versa, the nesting of union information terms must be described by the inverted weak synergy ordering. There is one union information for every antichain in the synergy lattice except for $\{\{\}\}$ which captures all information if the weak synergy is applied to it and hence captures no information if the union information is applied to it. Instead the antichain $\{1,\ldots,n\}$ is included because it captures no information with respect to weak synergy and hence all information with respect to union information. Thus the nesting of union information terms is described by the semi-lattice $(\mathcal{A},\succeq^\prime)$. It is not a full lattice because it has multiple lowest elements. See Figure \ref{fig:babel_to_boole:3_union_lattice} for the case $n=3$. The solution for the information atoms is not a Moebius-Inversion. The system of equations is still invertible because it is merely an equivalence transformation of the weak synergy system. Given a specific measure of union information $I_\cup^\ast$ the solution for the information atoms is equal to their solution in the weak synergy system where we set
\begin{equation}
I_{\ws}^\ast(\mathbf{a}_1,\ldots,\mathbf{a}_m:T) := I(\{1,\ldots,n\}:T) - I_{\cup}^\ast(\mathbf{a}_1,\ldots,\mathbf{a}_m:T)
\end{equation}
Union information is symmetric, subset-invariant, and satisfies
\begin{equation}
I_\cup(\mathbf{a}:T) = I(\mathbf{a}:T)
\end{equation}
The domain of the partner measure of union information $\overline{I_\cup}$ is the image of $\mathcal{A}$ under $\alpha \rightarrow \overline{\alpha}$, i.e. $\overline{A} = \mathbb{A} \backslash \{\{\{\}\}, \{\{1\},\{2\}\}\}$ which is not equal to any domain we have considered before. Since it is the complement of restricted information $I_\text{res}= \overline{I_\text{ws}}$, its nesting is described by the semi-lattice $(\overline{\mathcal{A}},\preceq)$. $\overline{I_\cup}$ is symmetric, superset-invariant and satisfies 
\begin{equation}
\overline{I_\cup}(\{i_1\},\ldots,\{i_m\}:T) = I(\{i_1,\ldots,i_m\}:T)
\end{equation}

\begin{figure}[ht] 
	\centering
	\includegraphics[width=0.75\textwidth]{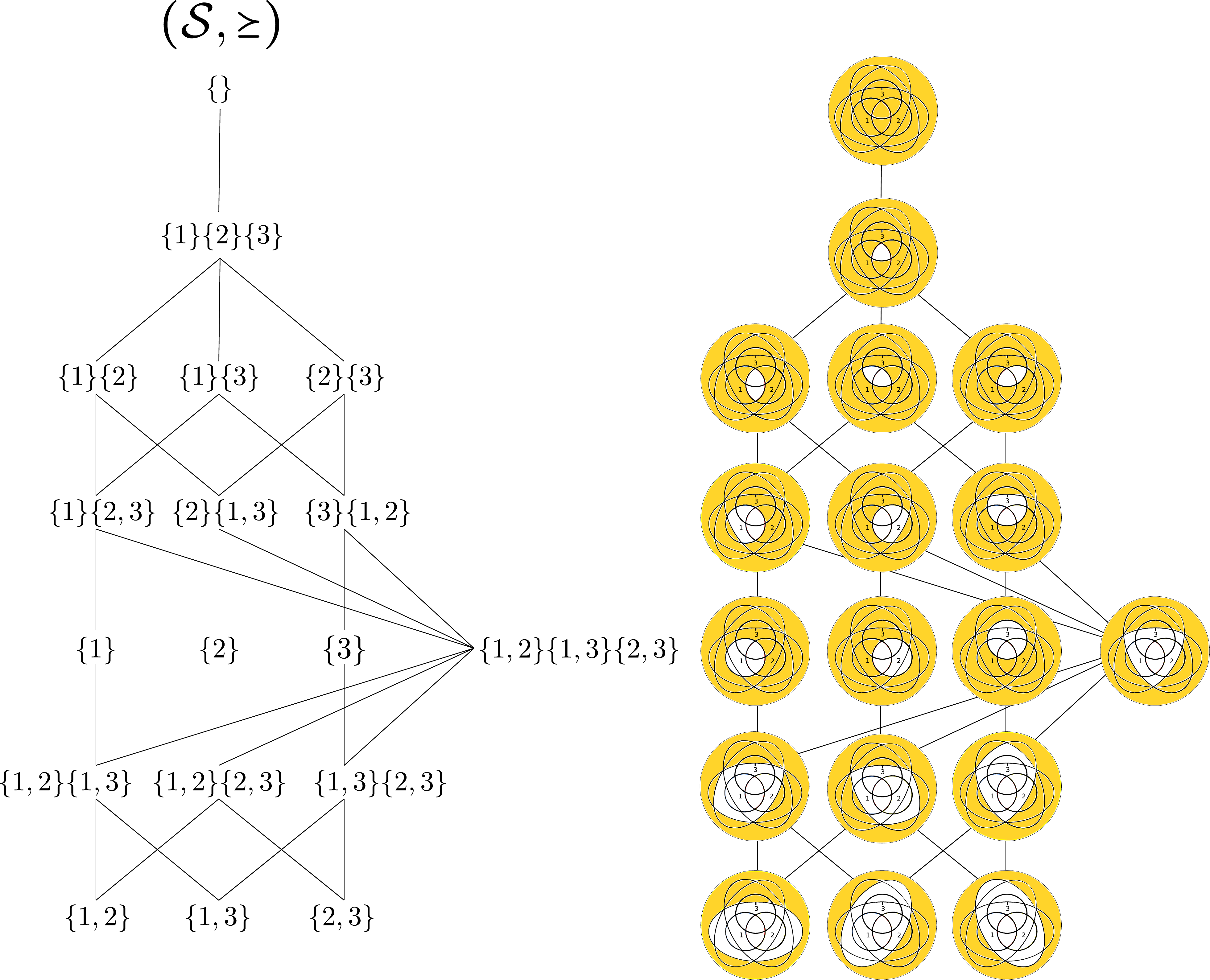}
	\caption{Left: Vulnerable information semi-lattice for $n=3$ sources. Right: Mereological information diagrams depicting the different vulnerable information terms.}
	\label{fig:babel_to_boole:vulnarable_3_lattice}
\end{figure}

\paragraph{Vulnerable information and its partner:}  Since vulnerable information is the complement of redundancy the nesting of vulnerable information terms must be described by the inverted redundancy ordering. Anologously to the the discussion of union information we conclude that the domain of vulnerable information is $\mathbb{A} \backslash \{\{\}, \{1,\ldots,n\}\} = \mathcal{S}$. Hence, the nesting of vulnerable information terms is described by the semi-lattice $(\mathcal{S},\succeq)$. Again, the solution for the information atoms does not have the structure of a Moebius-Inversion. See Figure \ref{fig:babel_to_boole:vulnarable_3_lattice} for the case $n=3$. The underlying system of equations is an equivalence transformation of the redundancy system and is therefore solvable. Given a specific measure of vulnerable information $I_\text{vul}^\ast$ the solution for the information atoms is equal to their solution in the redundancy system where we set
\begin{equation}
I_{\cap}^\ast(\mathbf{a}_1,\ldots,\mathbf{a}_m:T) := I(\{1,\ldots,n\}:T) - I_{\text{vul}}^\ast(\mathbf{a}_1,\ldots,\mathbf{a}_m:T)
\end{equation}
Vulnerable information is symmetric, superset-invariant, and satisfies
\begin{equation}
I_\text{vul}(\mathbf{a}:T) = I(\mathbf{a}^C:T|\mathbf{a})
\end{equation}
The partner of vulnerable information $\underline{I_\text{vul}}$ is defined on the domain 
\begin{equation}
\underline{\mathcal{S}} = \mathbb{A} \backslash \{\{\{1\},\ldots,\{n\}\}, \allowbreak \{\{1,\ldots,n\}\} \}
\end{equation}
which again is different from those we considered before. Since  $\underline{I_\text{vul}}$ is the complement of $\underline{I_\cap}$ its semi-lattice must be $(\underline{\mathcal{S}}, \preceq^\prime)$. $\underline{I_\text{vul}}$ is symmetric, subset-invariant and satisfies
\begin{equation}
\underline{I_\text{vul}}([n]\backslash\{i_1\},\ldots,[n]\backslash\{i_m\}:T) = I(\{i_1,\ldots,i_m\}^C:T|\{i_1,\ldots,i_m\})
\end{equation}

\paragraph{Inclusion-Exclusion} The logical conditions defining the different base-concepts do not only entail their individual properties as discussed above. Since each of them stands in an invertible relation to the information atoms, fixing one of them automatically fixes the others as well. We would like to illustrate this for the base-concepts of redundancy and union information. Based on their defining logical conditions on parthood relations these base-concepts must stand in an inclusion-exclusion relationship:
\footnotesize
\begin{align}
I_\cup(\mathbf{a}_1,\ldots,\mathbf{a}_m:T) &= \sum\limits_{1\leq i\leq m} \sum\limits_{f(\mathbf{a}_i)=1} \Pi(f) -  \sum\limits_{1\leq i < j \leq m} \sum\limits_{\substack{f(\mathbf{a}_i)=1 \\ f(\mathbf{a}_j)=1}  } \Pi(f) +  \sum\limits_{1\leq i < j < k  \leq m} \sum\limits_{\substack{f(\mathbf{a}_i)=1 \\ f(\mathbf{a}_j)=1 \\  f(\mathbf{a}_k)= 1}  } \Pi(f) - \ldots  \nonumber \\
&= \sum\limits_{1\leq i\leq m} I_\cap(\mathbf{a}_i:T) - \sum\limits_{1\leq i <  j\leq m} I_\cap(\mathbf{a}_i,\mathbf{a}_j:T) +\sum\limits_{1\leq i <  j< k\leq m} I_\cap(\mathbf{a}_i,\mathbf{a}_j,\mathbf{a}_k:T) - \ldots \nonumber
\end{align}
\normalsize
To see why the first equation is true consider its first summand. It involves all the information atoms that are part of at least one $I(\mathbf{a}_i:T)$. These are by construction exactly the atoms making up the union information $I_\cup(\mathbf{a}_1,\ldots,\mathbf{a}_m:T)$. However, some of these atoms are counted multiple times in the first summand. In particular, if such an atom is part of $k$ mutual information terms $I(\mathbf{a}_i:T)$, it will be counted $k$ times. So the remaining summands must make sure that each atom is only counted exactly once. This is true for the following reason: take any information atom $\Pi(f)$ appearing in the first summand and assume it is part of $k$ mutual information terms $I(\mathbf{a}_i:T)$. It is counted $k$ times by the first summand, $k \choose 2$ times by the second summand, $k \choose 3$ times by the third one, and so on until the $k$-th summand which counts it one time. So in total it is counted $\sum\limits_{i=1}^k (-1)^{i+1} {k \choose i} = 1$ times, as desired.

\FloatBarrier

\section{Relation to previous approaches} 
\label{sec:babel_to_boole:rel_prev_appr}

\subsection{Modified Synergistic Disclosure} \label{sec:babel_to_boole:modified_sd}
Rosas et al \cite{rosas2020operational} recently introduced a well motivated measure of synergistic information that is conceptually very similar to the notion of weak synergy introduced in the previous section. The measure is based on the idea of \textit{synergistic observables}.  Given an antichain $\alpha = \{\mathbf{a}_1,\ldots,\mathbf{a}_m\}$ an $\alpha$-synergistic observable $V$ is a univariate random variable such that $I(V:\mathbf{a}_i)=0$ for $i=1,\ldots,m$. In other words, a synergistic observable does not contain any information about an individual collection $\mathbf{a}_i$. The synergy of source collections $\mathbf{a}_1,\ldots,\mathbf{a}_m$ is then defined as the supremum of the information provided by synergistic observables that additionally satisfy the Markov condition $V-S-T$:
\begin{equation}
I_{SD}(\alpha:T) = \sup\limits_{\substack{\text{V is }\alpha-synergistic \\ V-S-T}} I(V:T)
\end{equation}
Intuitively, the Markov condition ensures that the information we are considering is actually contained in the sources so that, once we know them, V does not yield any additional information about the target. One may now introduce synergistic disclosure atoms via a Moebius inversion over the synergy lattice (or, as Rosas et al call it, the ``extended constraint lattice'')~\cite{rosas2020operational}. However, the resulting decomposition is not a standard PID because the \textit{consistency condition} \eqref{eq:babel_to_boole:consistency} is not satisfied. This means that the atoms cannot be interpreted in terms of parthood relations with respect to mutual information terms as described in Section \ref{sec:babel_to_boole:mereological_approach}. For example, we do not obtain any atoms interpretable as unique or redundant information in the case of two sources. This is because there are no two atoms in the decomposition that would necessarily add up to $I(S_1:T)$. But if there were atoms interpretable as the redundancy between the two sources and unique information of source 1 respectively, then these \textit{should always} add up to $I(S_1:T)$ (The same problem also arises for $I(S_2:T)$).

In order to construct a standard PID out of the synergistic disclosure measure one may however replace the self-disclosures with the appropriate conditional mutual information terms to enforce the consistency condition \eqref{eq:babel_to_boole:consistency} to be be satisfied. The resulting modified synergistic disclosure measure is defined as:
\begin{equation}
I_{MSD}(\alpha:T) = 
\begin{cases}
I_{SD}(\alpha:T) & \text{ if } |\alpha|\geq 2\\
I(T:\mathbf{a}_1^C|\mathbf{a}_1) & \text{ if } |\alpha| = 1.
\end{cases}
\end{equation}

\subsection{Loss and Gain Lattices}
In a 2017 paper Chicharro and Panzeri \cite{chicharro2017synergy} introduced partial information decompositions based on what they call information gain and information loss lattices. Structurally, these correspond to what we have here called redundancy lattices and synergy lattices, respectively. And in fact, there is an intuitive way to understand redundancy as an information gain and weak synergy as an information loss: suppose that initially we do not have access to any information source. Now we get access to at least one collection of sources $\mathbf{a}_1, \ldots, \mathbf{a}_m$ (we do not know which). Then the information that we are \textit{guaranteed to gain} should be exactly the redundancy between the $\mathbf{a}_i$. Hence, any redundancy can be described as the guaranteed information gain under such circumstances. Similarly, suppose that initially we have access to all sources. Now we lose access to all except one of the collections $\mathbf{a}_1, \ldots, \mathbf{a}_m$ (we do not know which). Then what we are left with will be exactly the information contained in the remaining collection (which could be any of them) and thus the information we are \textit{guaranteed to lose} should be exactly the information \textit{not} contained in any individual $\mathbf{a}_i$, i.e. their weak synergy. Hence, any weak synergy can be described as the guaranteed information loss under such circumstances. 

Indeed, the information loss decomposition is structurally identical to the weak synergy decomposition. It expresses a function of cumulative information loss as a downwards sum over the lattice $(\mathcal{S},\preceq^\prime)$. We would like to point out two differences between the construction of Chicharro and Panzeri and the one presented here: \textit{Firstly}, we start the construction with a characterization of the \textit{components} $\Pi$ of the mutual information decomposition. Composite information quantities such as redundancy or synergy are introduced via their appropriate relation to these components. The appropriate domains and lattices describing their nested structure can be \textit{derived} from these relations. By contrast the information gain (redundancy-based) and information loss (synergy-based) decompositions are introduced as two separate decompositions involving prima facie distinct sets of information atoms $\Delta I$ and $\Delta L$~\cite{chicharro2017synergy}. These are implicitly defined via Moebius-Inversion over the corresponding lattices. This raises the question of how these sets of components are to be interpreted and what their relation should be~\cite{chicharro2017synergy}. Due to the different construction these issues do not arise in the mereological approach. \textit{Secondly}, in the mereological approach redundancy- and synergy-based PID are just special cases of a more general unifying principle allowing the construction of information decompositions in terms of a great variety of base-concepts as discussed in Section~\ref{sec:babel_to_boole:logical_organization}. These base-concepts differ merely in their characteristic logical condition on parthood distributions.

\section{Conclusion} \label{sec:babel_to_boole:conclusion}
We presented a general pattern of logical conditions on parthood relations that captures all the PID base-concepts in the literature and that additionally leads to similarly interpretable novel base-concepts. These include in particular the concept of ``vulnerable information'', i.e.\ information we cannot obtain from at least one of the source collections at which it is evaluated. This concept may prove useful in a data security context where it the amount of information at risk of being lost since it is not entirely redundant. An interesting fact about vulnerable information is that its nested structure is described only by a semi-lattice and that its underlying system of equations does not have the structure of a Moebius-Inversion. This is how it differs from redundancy or weak synergy. Nonetheless its relationship to the information atoms is invertible and hence leads to a unique PID. The same applies to the concept of union information. 

Our construction also leads to ``partner measures'' for each of the PID base-concepts. These describe the same components of the joint mutual information but from the perspective of different antichains. Accordingly, two partner measures have different domains and (semi-)lattices describing their nested structure. One insight to be gained from this is that a synergy-based PID (in the form of its partner measure $\overline{I_{\text{ws}}}$) is obtainable via an upwards Moebius-Inversion on the redundancy lattice while a redundancy-based PID (in the form of its partner $\underline{I_\cap}$) can be obtained via an upwards Moebius-Inversion over the synergy-lattice. Overall, the unifying analysis presented here provides, on the one hand, more theoretical options for inducing PIDs that might be particularly suitable for certain applications contexts and, on the other, it lays the groundwork for detailed theoretical studies into the compatibility between properties of different base-concepts as functions of the underlying joint distribution. This latter point will be a particularly intriguing topic for future studies.

\section{Appendix}
\subsection{Proof that the partner measure mappings are inverses of each other} \label{sec:babel_to_boole:proof_partner_inverses}
First note that the non-subsets of the $\underline{\mathbf{a}} \in \underline{\alpha}$ are exactly the supersets of the $\mathbf{a} \in \alpha$, i.e. 
\begin{equation}
\{\mathbf{b} \in [n] : \neg \exists \underline{\mathbf{a}}\in \underline{\alpha}:\mathbf{b} \subseteq \underline{\mathbf{a}}\} = \{\mathbf{b} \in [n] : \exists \mathbf{a} \in \alpha: \mathbf{b} \supseteq \mathbf{a}\}
\end{equation}
Suppose $\mathbf{b} \in [n]$ is an element of the LHS so that $\neg \exists \underline{\mathbf{a}}\in \underline{\alpha}:\mathbf{b} \subseteq \underline{\mathbf{a}}$ and assume that it is not contained in the RHS so that $\mathbf{b}$ is a non-superset of the $\mathbf{a}\in \alpha$. But then $\mathbf{b}$ must be a subset of some $\underline{\mathbf{a}}\in \underline{\alpha}$ since these are the maximal non-supersets of the $\mathbf{a}\in \alpha$. This contradicts our initial assumption. Hence, if $\mathbf{b}$ is in the LHS it must be in the RHS. 

Now suppose that $\mathbf{b} \in [n]$ is an element of the RHS, i.e. it is a superset of some $\mathbf{a}\in \alpha$ and assume that it is not in the LHS because $\mathbf{b}$ is a subset of some $\underline{\mathbf{a}}\in \underline{\alpha}$. But then, since the $\underline{\mathbf{a}} \in \underline{\alpha}$ are the maximal non-supersets of the $\mathbf{a}\in \alpha$, $\mathbf{b}$ must be a non-superset of the $\mathbf{a}\in \alpha$ as well. Again this contradicts our initial assumption so that if $\mathbf{b}$ is in the RHS it must be in the LHS.

Therefore we have,

\begin{equation}
\overline{(\underline{\alpha})} = \min \{\mathbf{b} \in [n] : \neg \exists \underline{\mathbf{a}}\in \underline{\alpha}:\mathbf{b} \subseteq \underline{\mathbf{a}}\} = \min \{\mathbf{b} \in [n] : \exists \mathbf{a} \in \alpha: \mathbf{b} \supseteq \mathbf{a}\} = \alpha
\end{equation}

\bibliographystyle{unsrt}

\begin{thebibliography}{10}

\bibitem{williams2010nonnegative}
Paul~L Williams and Randall~D Beer.
\newblock Nonnegative decomposition of multivariate information.
\newblock {\em arXiv preprint arXiv:1004.2515}, 2010.

\bibitem{Harder2013}
Malte Harder, Christoph Salge, and Daniel Polani.
\newblock {Bivariate measure of redundant information}.
\newblock {\em Physical Review E - Statistical, Nonlinear, and Soft Matter
  Physics}, 87(1):12130, 2013.

\bibitem{Griffith2014_common_randomness}
Virgil Griffith, Edwin Chong, Ryan James, Christopher Ellison, and James
  Crutchfield.
\newblock {Intersection Information Based on Common Randomness}.
\newblock {\em Entropy}, 16(4):1985--2000, apr 2014.

\bibitem{Griffith_Ho2015}
Virgil Griffith and Tracey Ho.
\newblock {Quantifying redundant information in predicting a target random
  variable}.
\newblock {\em Entropy}, 17(7):4644--4653, 2015.

\bibitem{Barrett2015}
Adam~B Barrett.
\newblock {Exploration of synergistic and redundant information sharing in
  static and dynamical Gaussian systems}.
\newblock {\em Physical Review E}, 91(5):52802, 2015.

\bibitem{ince2017measuring}
Robin~AA Ince.
\newblock Measuring multivariate redundant information with pointwise common
  change in surprisal.
\newblock {\em Entropy}, 19(7):318, 2017.

\bibitem{finn2018pointwise}
Conor Finn and Joseph Lizier.
\newblock Pointwise partial information decomposition using the specificity and
  ambiguity lattices.
\newblock {\em Entropy}, 20(4):297, 2018.

\bibitem{makkeh2021isx}
Abdullah Makkeh, Aaron~J Gutknecht, and Michael Wibral.
\newblock Introducing a differentiable measure of pointwise shared information.
\newblock {\em Physical Review E}, 103(3):032149, 2021.

\bibitem{Kolchinsky2022}
Artemy Kolchinsky.
\newblock {A Novel Approach to the Partial Information Decomposition}.
\newblock {\em Entropy}, 24(3):403, 2022.

\bibitem{magri2021shared}
Cesare Magri.
\newblock On shared and multiple information.
\newblock {\em arXiv preprint arXiv:2107.11032}, 2021.

\bibitem{kleinman2021redundant}
Michael Kleinman, Alessandro Achille, Stefano Soatto, and Jonathan~C Kao.
\newblock Redundant information neural estimation.
\newblock {\em Entropy}, 23(7):922, 2021.

\bibitem{sigtermans2020pathbased}
David Sigtermans.
\newblock A path-based partial information decomposition.
\newblock {\em Entropy}, 22(9):952, 2020.

\bibitem{bertschinger2014quantifying}
Nils Bertschinger, Johannes Rauh, Eckehard Olbrich, J{\"u}rgen Jost, and Nihat
  Ay.
\newblock Quantifying unique information.
\newblock {\em Entropy}, 16(4):2161--2183, 2014.

\bibitem{pakman2021estimating}
Ari Pakman, Amin Nejatbakhsh, Dar Gilboa, Abdullah Makkeh, Luca Mazzucato,
  Michael Wibral, and Elad Schneidman.
\newblock Estimating the unique information of continuous variables.
\newblock {\em Advances in neural information processing systems},
  34:20295--20307, 2021.

\bibitem{james2018unique}
Ryan~G James, Jeffrey Emenheiser, and James~P Crutchfield.
\newblock Unique information and secret key agreement.
\newblock {\em Entropy}, 21(1):12, 2018.

\bibitem{rosas2020operational}
Fernando~E Rosas, Pedro~AM Mediano, Borzoo Rassouli, and Adam~B Barrett.
\newblock An operational information decomposition via synergistic disclosure.
\newblock {\em Journal of Physics A: Mathematical and Theoretical},
  53(48):485001, 2020.

\bibitem{van2023pooling}
Steven~J van Enk.
\newblock Pooling probability distributions and partial information
  decomposition.
\newblock {\em Physical Review E}, 107(5):054133, 2023.

\bibitem{Griffith2014}
Virgil Griffith and Christof Koch.
\newblock {Quantifying synergistic mutual information}.
\newblock In {\em Guided self-organization: inception}, pages 159--190.
  Springer, 2014.

\bibitem{gutknecht2021bits}
Aaron~J Gutknecht, Michael Wibral, and Abdullah Makkeh.
\newblock Bits and pieces: Understanding information decomposition from
  part-whole relationships and formal logic.
\newblock {\em Proceedings of the Royal Society A}, 477(2251):20210110, 2021.

\bibitem{Bertschinger2013_shared_info}
Nils Bertschinger, Johannes Rauh, Eckehard Olbrich, and J{\"{u}}rgen Jost.
\newblock {Shared information—New insights and problems in decomposing
  information in complex systems}.
\newblock In {\em Springer Proceedings in Complexity}, pages 251--269.
  Springer, 2013.

\bibitem{ehrlich2022partial}
David~A Ehrlich, Andreas~C Schneider, Michael Wibral, Viola Priesemann, and
  Abdullah Makkeh.
\newblock Partial information decomposition reveals the structure of neural
  representations.
\newblock {\em arXiv preprint arXiv:2209.10438}, 2022.

\bibitem{Crampton2000}
Jason Crampton and George Loizou.
\newblock Two partial orders on the set of antichains.
\newblock {\em Research note, September}, 2000.

\bibitem{chicharro2017synergy}
Daniel Chicharro and Stefano Panzeri.
\newblock Synergy and redundancy in dual decompositions of mutual information
  gain and information loss.
\newblock {\em Entropy}, 19(2):71, 2017.

\bibitem{perrone2016hierarchical}
Paolo Perrone and Nihat Ay.
\newblock Hierarchical quantification of synergy in channels.
\newblock {\em Frontiers in Robotics and AI}, 2:35, 2016.

\end{thebibliography}

\end{document}